\documentclass[12pt,a4paper]{article}
\usepackage{amssymb,enumitem,booktabs,subfig,xcolor,todonotes,microtype,setspace}
\usepackage[top=1.4in, bottom=1.4in, left=1.35in, right=1.35in]{geometry}
\usepackage{natbib}
\usepackage{url}
\usepackage[pdftex,colorlinks=true,hypertexnames=false]{hyperref}
\usepackage{amsmath}
\usepackage{amsfonts}
\usepackage{epsfig}
\usepackage{graphics}
\usepackage{palatino,eulervm}
\usepackage{graphicx}
\usepackage{lscape}
\usepackage{rotating}

\definecolor{darkblue}{rgb}{0,0,.6}
\hypersetup{citecolor=darkblue,linkcolor=darkblue,urlcolor=darkblue}
\allowdisplaybreaks[4]

\providecommand{\U}[1]{\protect\rule{.1in}{.1in}}

\graphicspath{{plots/}}

\usepackage{mathrsfs}
\usepackage{multirow}
\usepackage{arydshln}
\usepackage{amsthm,thmtools}
\declaretheorem{theorem}
\declaretheorem{lemma}
\makeatletter
\def\th@newremark{\th@remark\thm@headfont{\bfseries}}
\makeatletter
\theoremstyle{newremark}
\newtheorem{remark}{Remark}

\newtheorem{assumption}{Assumption}
\declaretheoremstyle[
  spaceabove=6pt, spacebelow=6pt,
  headfont=\bfseries,
  notefont=\mdseries, notebraces={(}{)},
bodyfont=\normalfont,
  postheadspace=0.5em,
]{mystyle}

\begin{document}
	
	\title{Dimension Reduction and MARS}
	
	\author{{\normalsize Yu Liu\thanks{Department of Statistics and Data Science, National University of Singapore, Singapore.},\ \ \ Degui Li\thanks{%
Department of Mathematics, University of York, UK.},\ \ \ Yingcun Xia%
\thanks{%
Department of Statistics and Data Science, National University of Singapore, Singapore, \url{staxyc@nus.edu.sg}}}\\
{\normalsize\em National University of Singapore and University of York}
}
\date{{\normalsize This version: \today}}

	\maketitle
	
	\begin{abstract}
		The multivariate adaptive regression spline (MARS) is one of the popular estimation methods for nonparametric multivariate regression. However, as MARS is based on marginal splines, to incorporate interactions of covariates, products of the marginal splines must be used, which often leads to an unmanageable number of basis functions when the order of interaction is high and results in low estimation efficiency. In this paper, we improve the performance of MARS by using linear combinations of the covariates which achieve sufficient dimension reduction. The special basis functions of MARS facilitate calculation of gradients of the regression function, and estimation of these linear combinations is obtained via  eigen-analysis of the outer-product of the gradients. Under some technical conditions, the consistency property is established for the proposed estimation method. Numerical studies including both simulation and empirical applications show its effectiveness in dimension reduction and improvement over MARS and other commonly-used nonparametric methods in regression estimation and prediction.
	\end{abstract}
	

\section{Introduction}\label{sec1}
Nonparametric estimation is an effective tool in statistics and machine learning to capture a flexible nonlinear relationship between the response and explanatory variables, relaxing pre-specified model structural assumptions required in parametric estimation methods.  However, extension of the nonparametric regression estimation to the setting with multivariate regressors needs to be handled with care, as the required number of observations (to achieve given estimation accuracy) increases exponentially as the dimension of covariates increases, resulting in the so-called ``curse of dimensionality" \citep[e.g.,][]{FG96}. To address this problem, we often have to restrict the class of multivariate regression functions so that only the lower dimensional nonparametric functions are to be estimated. Commonly-used function classes include additive models \citep{HT86}, varying-coefficient models \citep{HT93}, partially linear models \citep{EGRW86} and single-index models \citep{HHI93}.  However, these restricted nonparametric estimation methods may have unstable numerical performance in practical data analysis when the regression function class is misspecified.  Hence, it is imperative to develop a fully nonparametric multivariate estimation method that can reduce the curse of dimensionality but need no restriction on the class of regression functions.

In nonparametric estimation, the regression function is often approximated by a linear expansion of base functions \citep[e.g., Chapter 5 of][]{HTF09}. In the case of multivariate covariates, the required number of basis functions in the approximation may increase dramatically as the dimension of covariates increases. A commonly-used idea to design a feasible estimation algorithm is to control model complexity and thus limit the number of basis functions. This can be done by adaptively scanning the set of basis functions and selecting only those which contribute significantly to the model fitting. Among a long list of existing estimation algorithms, the multivariate adaptive regression spline \citep[MARS,][]{F91} is arguably the most popular one. It uses piecewise linear basis functions and can be viewed as a natural generalization of the stepwise linear regression approach. Because of the selection of splines in the estimation algorithm, MARS can also result in variable selection. MARS is well suited for high-dimensional nonparametric regression problems and can be further extended to tackle classification problems \citep[e.g.,][]{SHKT97}. Existing literature in statistical learning such as \cite{HTF09} usually implements the MARS algorithm directly without making any transformation or dimension reduction of the covariates. This may result in an unmanageable number of basis functions (if the level of model complexity or the order of interaction is high) and low estimation efficiency.

In multivariate nonparametric regression, it is often the case that important features of multiple regressors are retrievable via low-dimensional projections. The low-dimensional sub-space is expected to retain all (or most of) the information provided by the covariates on the response, and is thus called the sufficient dimension reduction (SDR) space, which is first introduced by \cite{L91}. Aiming at dimension reduction for the conditional mean, which is more relevant to our main interest, a similar concept (central mean space) is also introduced by \cite{cook2002dimension}. More recent developments on this topic can be found in \cite{X08}, \cite{chen2010coordinate}, \cite{YL11}, \cite{FL14}, \cite{luo2014efficient}, \cite{Ma14}, \cite{WXZ15},  \cite{YBL17}, and \cite{fertl22}. This paper aims to combine SDR  with MARS by incorporating linear combinations of covariates to improve the regression estimation. These linear combinations are the SDR directions or more precisely the central mean space of \cite{cook2002dimension} when the underlying model has a multiple-index structure and can effectively reduce the order of covariate interaction required in MARS and improve the estimation performance. As these linear combinations in MARS are dimension-reduced covariates, the proposed methodology is called drMARS throughout the paper.

The nonparametric estimation procedure developed in this paper includes two stages: (i) estimate the SDR space of the conditional mean; and (ii) modify MARS by incorporating these linear combinations of covariates (or SDR) to estimate the regression functions. The main technique in stage (i) is to conduct eigen-analysis of the outer-product of regression function gradient estimates and estimate the SDR directions by the eigenvectors corresponding to the first few largest eigenvalues. In particular, we estimate the gradient via a linear basis expansion determined by MARS and further derive a sensible convergence property for the resulting estimates. With the MARS algorithm, this new gradient estimation is easy to implement, complementing other gradient estimation methods such as the local linear smoothing and reproducing kernel Hilbert space which have been extensively studied in the literature \citep[e.g.,][]{XTLZ02, X08, FL14}. The drMARS in stage (ii) incorporates the linear combinations of covariates, making it substantially different from the classic MARS in \cite{F91}. In particular, when a high-order interaction of covariates can be equivalently expressed as the multiple-index form, our drMARS can significantly reduce the number of terms in the basis expansion and improve the estimation efficiency. As a simple example, $(x_1 + x_2 + x_3 + x_4)^3$ has a third-order interaction when the conventional MARS is applied, but it has only a first-order interaction in the drMARS if the linear combination is correctly identified. This is confirmed by our numerical studies, which also show the advantage of drMARS even if the postulated model cannot reduce the order of interactions via the SDR-determined linear combinations of covariates. drMARS inherits some nice features from MARS (such as the simple form of linear spline basis functions and selection of spline in the algorithm) and works well when the dimension of predictors is relatively large (see the simulation and empirical application). Under some technical conditions, we derive the consistency theory for the drMARS estimation, complementing the existing asymptotic theory for the spline-based estimation \citep[e.g.,][]{S90, S91, ZSW98, H03, L13}.

Another work related to our approach is the random projection or random rotation \citep[e.g.,][]{BlaserFryzlewicz2016,cannings2017random,bagnall2020rotation}. The random rotation is an ensemble procedure. It randomly selects the projections and estimates the model using the projected  combinations of the variables as predictors for regression methods such as the random forest or support vector machine. Each set of projections thus generates a prediction. The final prediction is a weighted average of these predictions. In contrast, the rotation in our approach is based on the regression itself, i.e., SDR, and thus is more efficient for prediction. As we will show in the numerical studies, the rotation based on SDR has better estimation and prediction accuracy than the random rotation.

The rest of the paper is organized as follows. Section \ref{sec2} defines the SDR space, introduces the MARS-based estimation method, and develops the convergence properties of the estimates. Section \ref{sec3} describes the drMARS algorithm and its consistency theory. Sections \ref{sec4} and \ref{sec5} report the simulation studies and real data applications, respectively. Section \ref{sec6} concludes the paper. Proofs of the main theorems are available in an appendix. Throughout the paper, for a vector $u=(u_1,\cdots,u_d)^{^\intercal}$, we define $|u|_q^q=\sum_{i=1}^d |u_i|^q$ with $q\geq1$; for a $d\times d$ matrix ${\mathbf W}=(w_{ij})_{d\times d}$, we let $\Vert {\mathbf W}\Vert$ and $\Vert {\mathbf W}\Vert_F$ be the spectral and Frobenius norms, respectively.


\section{Estimation of SDR space via MARS}\label{sec2}

Let $Y$ and $X$ be the response and $p$-dimensional vector of covariates, respectively. Assume the following multiple-index model structure:
\begin{equation}\label{DR}
	G(x) = {\sf E}(Y | X=x) = {\sf E}\left(Y | {\mathbf B}^{^\intercal} X={\mathbf B}^{^\intercal} x\right)=G_0\left({\mathbf B}^{^\intercal} x\right),
\end{equation}
where ${\mathbf B}$ is a $p\times d$ orthogonal matrix with $d$ smaller than $p$, $G(\cdot)$ is a multivariate nonparametric regression function on ${\cal R}^p$ and $G_0(\cdot)$ is a nonparametric link function on ${\cal R}^d$. It follows from model (\ref{DR}) that projection of the $p$-dimensional $X$ onto the $d$-dimensional sub-space ${\mathbf B}^{^\intercal} X$ retains all the information provided by $X$ for prediction of $Y$. Hence, the matrix ${\mathbf B}$ determines the SDR directions (or the central mean subspace). The space spanned by ${\mathbf B}$'s column vectors is called the SDR space.

Letting $u={\mathbf B}^{^\intercal} x$, by (\ref{DR}), we readily have that $G^\prime(x)={\mathbf B}G_0^\prime(u)$, where $G^\prime$ and $G_0^\prime$ are the gradient vectors. By Lemma 1 in \cite{XTLZ02}, the space spanned by ${\mathbf B}$ is the same as that spanned by the eigenvectors of ${\boldsymbol\Sigma}_G:={\sf E}\left[G^\prime(X)G^\prime(X)^{^\intercal}\right]$ corresponding to the largest $d$ eigenvalues, i.e., ${\sf span}({\mathbf B})={\sf span}(\beta_1,\cdots,\beta_d)$, where $\beta_j$ is the eigenvector of ${\boldsymbol\Sigma}_G$ corresponding to the $j$-th largest eigenvalue. With a sample of observations $(Y_i, X_i)$, $i=1,\cdots,n$, we estimate ${\boldsymbol\Sigma}_G$ by the outer-product of gradient estimates:
\begin{equation}\label{OPG}
	\widetilde{\boldsymbol\Sigma}_G=\frac{1}{n} \sum_{i=1}^n\widetilde{G}^\prime(X_i)\widetilde{G}^\prime(X_i)^{^\intercal},
\end{equation}
where $\widetilde{G}^\prime$ is a nonparametric estimate of the gradient $G^\prime$. A natural estimate of $G^\prime$ is via the local linear smoothing method \citep[e.g.,][]{FG96}. The estimate of ${\mathbf B}$ can be obtained by subsequently conducting the eigen-analysis of $\widetilde{\boldsymbol\Sigma}_G$ \citep[e.g.,][]{XTLZ02, X08}. However, the local linear estimation is essentially a kernel-based local smoothing method which is sensitive to the smoothing parameter choice, and still suffers the ``curse of dimensionality'' when the dimension $ p $ is large.

Next, we propose an alternative technique to estimate $G^\prime$ via MARS. MARS is an adaptive estimation procedure using linear spline functions in the basis expansion. For the $k$-th covariate, we define the piecewise linear basis functions with knots taken from the set $\left\{t_{k,1},\cdots,t_{k,n_k}\right\}$:
\begin{eqnarray}
	h^+_{k,j}(x_k)&=&(x_k-t_{k,j})_{+}=\left\{\begin{array}{cc}
		x_k-t_{k,j}, & \text { if } x_k>t_{k,j}, \\
		0, & \text { otherwise },
	\end{array}  \quad\right.\label{basis+}\\
	h^-_{k,j}(x_k)&=&(x_k-t_{k,j})_{-}=(t_{k,j}-x_k)_{+},\label{basis-}
\end{eqnarray}
which form reflected pairs for the $k$-th covariate at $t_{k,j}$, $j=1,\cdots,n_k$. The collection of marginal basis functions for all the covariates is
$$
\mathcal{C}=\left\{\left(h^+_{k,j}, h^-_{k,j}\right),\ j=1,\cdots,n_k,\  k=1,2, \cdots, p\right\}.
$$
When each basis function depends only on a single covariate, the number of basis functions in ${\cal C}$ is $2\sum_{k=1}^p n_k$, assuming all the knots are distinct. To incorporate interactions of covariates, we use tensor products of the basis functions in ${\cal C}$. Specifically, when the order of interaction is set to be $R$, a typical $R$-variate basis function is defined as
\begin{equation}\label{R-basis}
	h_{k_1j_1,\cdots,k_Rj_R}(x_{k_1},\cdots,x_{k_R})=\prod_{r=1}^R h_{k_r,j_r}(x_{k_r}),
\end{equation}
where $h_{k,j}$ is a basis function from ${\cal C}$, $1\leq j_{r}\leq 2n_{k_r}$, and $1\leq k_1\neq k_2\neq\cdots\neq k_R\leq p$. Note that the number of the $R$-variate basis functions increases dramatically as $p$ increases.

Suppose that the multivariate nonparametric regression function is approximated by the following form of basis expansion:
\begin{equation}\label{basis-ex}
	G(x)\approx G_m(x):=\theta_{0}+\sum_{j=1}^{m} \theta_{j} h_{j}(x),
\end{equation}
where $h_j$ is either a basis function in ${\cal C}$ or a product of marginal basis functions, see (\ref{R-basis}), and $m$ is the number of basis functions which may diverge to infinity. Here $G(x)\approx G_m(x)$ means that $G_m(x)\rightarrow G(x)$ as $m\rightarrow\infty$. The coefficients $\theta_{j}$, $j=0,1,\cdots,m$, are estimated by the least squares as in standard linear regression. From (\ref{basis-ex}), we further obtain the basis expansion for the gradient $G^\prime$:
\begin{equation}\label{basis-gra}
	G^\prime(x)\approx G_m^\prime(x):=\sum_{j=1}^{m} \theta_{j} h_{j}^\prime(x),
\end{equation}
where $h_{j}^\prime$ is the gradient vector of $h_j$. When the order of interaction $R$ is large (or even moderately large), it is practically infeasible to include all the $R$-variate basis functions. The real art of MARS is to provide an adaptive selection procedure including both the forward and backward stepwise algorithms to construct the basis functions with the linear spline functions in ${\cal C}$. This adaptive selection reduces the number of basis functions in (\ref{basis-ex}) while retains the model flexibility.

We next briefly describe the MARS algorithm to determine the basis functions in (\ref{basis-ex}) and (\ref{basis-gra}). Start with the constant function $h_0(x)\equiv 1$ and use the linear spline functions in ${\cal C}$ as the candidate functions. In each stage, let ${\cal M}$ be the set of basis functions which have been selected in the previous stages. Construct a new basis function from products of any basis function in ${\cal M}$ with one of the reflected pairs in ${\cal C}$. This new term in the basis expansion has the following typical form:
\[
\theta_{|{\cal M}|+1}h_l(x)h^+_{k,j}(x_k)+ \theta_{|{\cal M}|+2}h_l(x)h^-_{k,j}(x_k),\ \ h_l\in{\cal M},\ \ \left(h^+_{k,j}, h^-_{k,j}\right)\in{\cal C},
\]
where $\theta_{|{\cal M}|+1}$ and $\theta_{|{\cal M}|+2}$ are the parameters to be estimated by least squares, and $|{\cal M}|$ denotes the cardinality of ${\cal M}$. Add the products to the model approximation with the basis functions in ${\cal M}$ and choose the product which results in the largest decrease in the training estimation errors. Repeat the above process until the number of the selected basis functions reaches a pre-determined number $M$. As the number $M$ is usually large, the model selected in the forward stepwise algorithm often overfits the data. Thus, a backward stepwise algorithm is needed to delete the term whose removal results in the smallest increase in the residual squared errors.

Let $\tilde h_1,\cdots,\tilde h_{\widetilde m}$ be the basis functions selected by MARS and $\widetilde h_{j}^\prime$ be the gradient vector of $\widetilde h_j$, $j=1,\cdots,\widetilde{m}$. We write
\[
	\widetilde{\mathbf H}(x)=\left[1,\widetilde h_1(x),\cdots, \widetilde h_{\widetilde m}(x)\right]^{^\intercal}\ \ {\rm and}\ \ \widetilde{\mathbf H}^\prime(x)=\left[{\mathbf 0}_p, \widetilde h_1^\prime(x),\cdots,\widetilde h_{\widetilde m}^\prime(x)\right]^{^\intercal},
\]
where ${\mathbf 0}_p$ is a $p$-dimensional zero vector. We use least squares to estimate the parameters in the final basis approximation:
\begin{equation}\label{est-LS}
	\widetilde{\boldsymbol\alpha}=\left(\widetilde{\alpha}_0,\widetilde{\alpha}_1,\cdots,\widetilde{\alpha}_{\widetilde m}\right)^{^\intercal}=\left(\widetilde{\mathbb H}^{^\intercal}\widetilde{\mathbb H}\right)^{-1}\widetilde{\mathbb H}^{^\intercal} {\mathbf Y},
\end{equation}
where 
\[
\widetilde{\mathbb H}=\left[\widetilde{\mathbf H}(X_1),\cdots,\widetilde{\mathbf H}(X_n)\right]^{^\intercal}\ \ {\rm and}\ \ {\mathbf Y}=(Y_1,\cdots,Y_n)^{^\intercal}.
\]
Consequently the estimate of $G^\prime(x)$ can be obtained by
\begin{equation}\label{est-1}
	\widetilde{G}^\prime(x)=\sum_{j=1}^{\widetilde m} \widetilde\alpha_j\widetilde h_j^\prime(x)=\widetilde{\mathbf H}^\prime(x)^{^\intercal}\left(\widetilde{\mathbb H}^{^\intercal}\widetilde{\mathbb H}\right)^{-1}\widetilde{\mathbb H}^{^\intercal} {\mathbf Y}.
\end{equation}
The above gradient estimate is then used to construct $\widetilde{\boldsymbol\Sigma}_G$ in (\ref{OPG}). Letting $\widetilde\beta_j$ be the eigenvector of $\widetilde{\boldsymbol\Sigma}_G$ corresponding to the $j$-th largest eigenvalue, we obtain $\widetilde{\mathbf B}=(\widetilde\beta_1,\cdots,\widetilde\beta_d)$, which will be shown to be a consistent estimate of ${\mathbf B}$ (subject to appropriate rotation); see Theorem \ref{th:2} below.

If the order of covariate interaction is set as $R$ in MARS, we let ${\mathbf H}(\cdot)$ be a vector containing all the basis functions which are either from ${\mathcal C}$ or tensor products of the marginal basis functions as in (\ref{R-basis}). Without loss of generality, we may write 
\[{\mathbf H}(\cdot)=\left[\widetilde{\mathbf H}(\cdot)^{^\intercal}, \widetilde {\mathbf H}_{-}(\cdot)^{^\intercal}\right]^{^\intercal},\]
where $\widetilde {\mathbf H}_{-}(\cdot)$ is a vector of basis functions not selected by MARS. Let $m_H$ be the dimension of ${\mathbf H}(\cdot)$ which is often much larger than $\widetilde m$. It is worth pointing out that ${\mathbf H}(\cdot)$ is a vector of deterministic functions which can be seen as the candidate basis functions in MARS, and $m_H$ is a non-random positive integer.

We next study the convergence property for the MARS-based nonparametric estimates $\widetilde{G}^\prime$ and $\widetilde{\boldsymbol\Sigma}_G$, which requires the following technical conditions.

\begin{assumption}\label{ass:1}
	
	{\em (i)\ Let $(Y_i,X_i)$, $i=1,\cdots,n$, be independent and identically distributed (i.i.d.), and $\varepsilon_i:=Y_i-G(X_i)$ be zero-mean and homoskedastic, i.e.,  ${\sf E}(\varepsilon_i^2|X_i)=\sigma^2>0$ almost surely (a.s.). 
	
	(ii)\ The density function of $X_i$ exists, and is bounded away from zero and infinity on a compact set ${\cal X}$. Both $G$ and $G^\prime$ are continuous on ${\cal X}$.
	
	(iii)\ The matrix ${\boldsymbol\Omega}:={\sf E}\left[{\mathbf H}(X_i) {\mathbf H}(X_i)^{^\intercal}\right]$ is positive definite, and $m_H\sqrt{\log m_H}=o(n)$.
	
	(iv)\ There exists $\widetilde\rho(\cdot)$ satisfying $\widetilde\rho(u)\rightarrow0$ as $u\rightarrow\infty$, such that 
 \[\sup_{x\in{\cal X}}\left\vert G^\prime(x)-\widetilde{\mathbf H}^\prime(x)^{^\intercal}{\boldsymbol\alpha}_o\right\vert_2=O_P\left(\widetilde\rho(m_\ast)\right),\ \ {\boldsymbol\alpha}_o=\left(\widetilde{\mathbb H}^{^\intercal} \widetilde{\mathbb H}\right)^{-1}\widetilde{\mathbb H}^{^\intercal}{\mathbf G},\]
 conditional on $\widetilde m=m_\ast$, where ${\mathbf G}=\left[G(X_1),\cdots,G(X_n)\right]^{^\intercal}$. 

	(v)\ The matrix ${\boldsymbol\Sigma}_G$ has full rank of $d$ with positive and distinct eigenvalues.}
	
\end{assumption}

\begin{remark}\label{re:1}
	
	The independence restriction in Assumption \ref{ass:1}(i) can be weakened and the theory developed in this section also holds for stationary and weakly dependent time series satisfying some mixing properties \citep[e.g.,][]{B05}. Assumption \ref{ass:1}(ii) is commonly used in deriving asymptotic results of the spline-based estimation. The compact support restriction can be relaxed at the cost of slightly more lengthy proof with some moment condition on $X$.
	
	Assumption \ref{ass:1}(iii) is a sufficient condition to ensure that the least squares estimate in (\ref{est-LS}) is well defined. In fact, by Lemma \ref{le:2} in the appendix, $\frac{1}{n}\widetilde{\mathbb H}^{^\intercal}\widetilde{\mathbb H}$ is positive definite with probability approaching one (w.p.a.1), which is implicitly assumed in \cite{F91}. As the linear spline basis functions are special polynomial spline functions, we may replace Assumption \ref{ass:1}(iii) by an alternative condition through \cite{H03}'s theoretical framework. Let $\widetilde{\cal S}$ be the estimation space containing the linear spline functions and their tensor products selected by MARS. Given a sample of covariates $X_1,\cdots,X_n$, suppose that $\widetilde{\cal S}$ is empirically identifiable in the sense that $g\in\widetilde{\cal S}$  and $\vert g\vert_n^2=\frac{1}{n}\sum_{i=1}^ng^2(X_i)=0$ together imply $g\equiv0$. For a $p$-dimensional vector ${\mathbf v}$, ${\mathbf v}^{^\intercal}( \frac{1}{n}\widetilde{\mathbb H}^{^\intercal}\widetilde{\mathbb H}){\mathbf v}=0$ indicates that $\vert {\mathbf v}^{^\intercal}\widetilde{\mathbf H}\vert_n^2=\frac{1}{n}\sum_{i=1}^n[{\mathbf v}^{^\intercal}\widetilde{\mathbf H}(X_i)]^2=0$. Then, as ${\mathbf v}^{^\intercal}\widetilde{\mathbf H}\in\widetilde{\cal S}$, by the empirical identifiability of $\widetilde{\cal S}$, we readily have ${\mathbf v}^{^\intercal}\widetilde{\mathbf H}(x)=0$ for any $x\in{\cal X}$, and thus ${\mathbf v}=0$. This shows that $\frac{1}{n}\widetilde{\mathbb H}^{^\intercal}\widetilde{\mathbb H}$ is positive definite w.p.a.1, and its inverse is well defined. Assumption \ref{ass:1}(iii) restricts the divergence rate of $m_H$, which is very mild for the nonparametric series estimation.

	
	Assumption \ref{ass:1}(iv) imposes a high-level condition on the uniform bias order of the gradient estimate. In fact, ${\boldsymbol\alpha}_o^{^\intercal}\widetilde{\mathbf H}(\cdot)$ can be seen as the projection of $G$ onto the estimation space $\widetilde{\cal S}$ defined above. In the spline-based estimation theory, it is reasonable to assume $\vert G(x)-{\boldsymbol\alpha}_o^{^\intercal}\widetilde{\mathbf H}(x)\vert\rightarrow 0$ uniformly over $x\in{\cal X}$. Assumption \ref{ass:1}(iv) shows that this uniform approximation continues to hold when $G$ and its projection onto $\widetilde{\cal S}$ are replaced by their gradients. Let ${\cal S}$ be the estimation space containing all the linear spline functions and their tensor products as in ${\mathbf H}(\cdot)$. It is clear that $\widetilde{\cal S}\subset{\cal S}$. Letting ${\mathbb H}=\left[{\mathbf H}(X_1),\cdots, {\mathbf H}(X_n)\right]^{^\intercal}$, we define ${\boldsymbol\alpha}_\dagger=({\mathbb H}^{^\intercal}{\mathbb H})^{-1}{\mathbb H}^{^\intercal}{\mathbf G}$ so that ${\boldsymbol\alpha}_\dagger^{^\intercal}{\mathbf H}(\cdot)$ can be seen as the projection of $G$ onto ${\cal S}$. The bias term of the gradient estimation can be decomposed as 
 \begin{eqnarray}
        G^\prime(x)-\widetilde{\mathbf H}^\prime(x)^{^\intercal}{\boldsymbol\alpha}_o&=&\left[G^\prime(x)-{\mathbf H}^\prime(x)^{^\intercal}{\boldsymbol\alpha}_\dagger\right]+\left[{\mathbf H}^\prime(x)^{^\intercal}{\boldsymbol\alpha}_\dagger-\widetilde{\mathbf H}^\prime(x)^{^\intercal}{\boldsymbol\alpha}_o\right]\notag\\
        &=:&\widetilde{\rm bias}_1\left(x\right)+\widetilde{\rm bias}_2\left(x\right).\label{bias-1}
 \end{eqnarray}
The first term $\widetilde{\rm bias}_1\left(x\right)$ is due to the approximation error of $G^\prime(\cdot)$ by its projection onto the space ${\cal S}$. In fact, under some smoothness conditions on $G$ and $G^\prime$ \citep[e.g.,][]{S82, H03}, with the approximation theory, we conjecture that its order is upper bounded by $m_H^{-q/p}$, where $q$ is a positive number relevant to the order of bounded and continuous derivatives of $G(\cdot)$. The second term $\widetilde{\rm bias}_2\left(x\right)$ is induced by the projection onto the MARS-selected estimation space $\widetilde{\cal S}$ rather than ${\cal S}$. According to the MARS algorithm, we expect this bias order tends to zero if $\widetilde{m}$ is sufficiently large. If $\widetilde m$ is of the same order as $m_H$, it is reasonable to conjecture that the two terms on the right side of (\ref{bias-1}) have the same approximation order. 
	
	Assumption \ref{ass:1}(v) is analogous to the condition (C4) in \cite{X08}, making it feasible to apply the Davis-Kahan theorem \citep[e.g., Theorem 2 in][]{YWS15} to prove Theorem \ref{th:2}.
	
\end{remark}


\begin{theorem}\label{th:1}
	
	Suppose that Assumption \ref{ass:1}(i)--(iv) is satisfied. Then, conditional on $\widetilde m=m_\ast$,
	\begin{eqnarray}
		&&\left\vert\widetilde{G}^\prime(x)-G^\prime(x)\right\vert_2=O_P\left( m_\ast^{1/2}n^{-1/2}+\widetilde\rho(m_\ast)\right),\label{th1-1}\\
		&&\left\Vert\widetilde{\boldsymbol\Sigma}_G-{\boldsymbol\Sigma}_G\right\Vert=O_P\left( m_\ast^{1/2}n^{-1/2}+\widetilde\rho(m_\ast)\right).\label{th1-2}
	\end{eqnarray}
	
\end{theorem}


\begin{remark}\label{re:2}
	
	The two convergence rates in (\ref{th1-1}) and (\ref{th1-2}) are due to the estimation variance and bias, respectively. They are comparable to the convergence results obtained by \cite{FL14}, where the gradient is estimated by the covariance operator on the reproducing kernel Hilbert space. It is worth noting that the rates in (\ref{th1-1}) and (\ref{th1-2}) are slower than the root-$n$ rate as MARS is essentially nonparametric. Furthermore, if the minimum eigenvalue of ${\boldsymbol\Omega}$ converges slowly to zero, the convergence rates would be further slowed down. For instance, we may show the following convergence property for the MARS estimate of the gradient: 
 \[\left\vert\widetilde{G}^\prime(x)-G^\prime(x)\right\vert_2=O_P\left(\left( m_\ast/\underline\lambda\right)^{1/2}n^{-1/2}+\widetilde\rho(m_\ast)\right),\ \  \underline\lambda=\lambda_{\min}({\boldsymbol\Omega}),
 \]
 conditional on $\widetilde m=m_\ast$.
	
	In Theorem \ref{th:1}, we assume that the number of candidate covariates in nonparametric regression is fixed. The convergence results in (\ref{th1-1}) and (\ref{th1-2}) can be further extended to the setting when the covariate number is divergent at a slow polynomial rate of $n$. Following the proof of Theorem \ref{th:1} in the appendix and assuming that $\widetilde S$ is empirically identifiable, we may show that
	\begin{eqnarray}
		&&\left\vert\widetilde{G}^\prime(x)-G^\prime(x)\right\vert_2=O_P\left(\left(p m_\ast\right)^{1/2}n^{-1/2}+\widetilde\rho(m_\ast)\right),\notag\\
		&&\left\Vert\widetilde{\boldsymbol\Sigma}_G-{\boldsymbol\Sigma}_G\right\Vert=O_P\left(p^{1/2}\left[\left(pm_\ast\right)^{1/2}n^{-1/2}+\widetilde\rho(m_\ast)\right]\right),\notag
	\end{eqnarray}
	conditional on $\widetilde m=m_\ast$. These convergence properties indicate that the dimension $p$ must be of order smaller than $n^{1/2}$. For  high-dimensional nonparametric estimation with $p$ possibly larger than $n^{1/2}$, we may have to impose sparsity assumptions on $G^\prime$ and ${\boldsymbol\Sigma}_G$, and combine the developed MARS estimation with a shrinkage technique \citep[e.g.,][]{BL08}.

\end{remark}


\begin{theorem}\label{th:2}
	
	Suppose that Assumption \ref{ass:1}(i)--(v) is satisfied. Conditional on $\widetilde m=m_\ast$, there exists a $d\times d$ rotation matrix ${\mathbf Q}$ such that $\left\Vert\widetilde{\mathbf B}-{\mathbf B}{\mathbf Q}\right\Vert=O_P\left( m_\ast^{1/2}n^{-1/2}+\widetilde\rho(m_\ast)\right)$.
	
\end{theorem}


\begin{remark}\label{re:3}
	
	\cite{X08} derives a faster convergence rate by using the minimum average variance estimation with refined kernel weights. However, some restrictive conditions are imposed on the smoothing parameter and the dimension $d$. For instance, $d$ cannot exceed $3$ to achieve the root-$n$ convergence. In contrast, we do not require additional restriction on $d$. 
	
	We need to determine the dimension of the SDR space for which many criteria have been proposed \citep[e.g.,][]{L91,XTLZ02}. In the simulation study, we select the dimension via the $10$-fold cross-validation (CV) criterion. We do not study the theory of the dimension selection in this paper, but our simulations suggest that this criterion works reasonably well; see Table \ref{dSDR}.
	
	 As the basis of the SDR space is not unique, the MARS estimate $\widetilde{\mathbf B}$ converges to ${\mathbf B}$ up to appropriate transformation via the rotation matrix ${\mathbf Q}$. However, with Assumption \ref{ass:1}(v) and the model identification conditions as in Proposition 1.1 of \cite{X08}, we may consider ${\mathbf Q}$ as an identity matrix and thus $\widetilde{\mathbf B}$ converges to ${\mathbf B}$. Since $ {\mathbf B}{\mathbf Q} $ is also a base of SRD space, for notational convenience, we do not distinguish between $ {\mathbf B} $ and ${\mathbf B}{\mathbf Q}$ in the rest of the paper, and use $ {\mathbf B} $ to denote both cases.
	
\end{remark}	


\section{Dimension-reduced MARS}\label{sec3}

Let
\[X_i^\ast=\widetilde{\mathbf B}^{^\intercal} X_i=\left(\widetilde{\beta}_1^{^\intercal} X_i,\cdots,\widetilde{\beta}_d^{^\intercal} X_i\right)^{^\intercal}\]
be a $d$-dimensional vector of projected covariates, where $\widetilde{\mathbf B}$ is defined in Section \ref{sec2}. Generally, we can use any SDR method, such as SIR of \cite{L91}, to estimate $ \widetilde{\mathbf B} $, and then apply MARS to $(Y_i, X_i^\ast)$, $i=1,\cdots,n$. We call this general approach SDR-MARS, while the MARS estimation based on our dimension reduction proposed in Section \ref{sec2} is still called drMARS to avoid possible confusion. Due to the convergence property of $\widetilde{\mathbf B}$ in Theorem \ref{th:2}, we expect that $X_i^\ast$ can well approximate $X_i^\circ={\mathbf B}^{^\intercal} X_i$. Write $x_\ast=\widetilde{\mathbf B}^{^\intercal} x=\left(x_1^\ast,\cdots,x_d^\ast\right)^{^\intercal}$ and $x_\circ={\mathbf B}^{^\intercal} x$, $x\in{\cal R}^p$. For the $k$-th projected covariate, we define $h^+_{k,j}(x_k^\ast)$ and $h^-_{k,j}(x_k^\ast)$ similarly to $h^+_{k,j}(x_k)$ and $h^-_{k,j}(x_k)$ in (\ref{basis+}) and (\ref{basis-}) but with the set of knots $\left\{t_{k,1},\cdots,t_{k,n_k}\right\}$ replaced by $\left\{t_{k,1}^\ast,\cdots,t_{k,n_k^\ast}^\ast\right\}$, and construct
\[
\mathcal{C}^\ast=\left\{\left(h^+_{k,j}, h^-_{k,j}\right),\ j=1,\cdots,n_k^\ast,\  k=1,2, \cdots, d\right\}.
\]
With a sample of response and projected covariates $(Y_1,X_1^\ast),\cdots,(Y_n,X_n^\ast)$, we use the linear spline functions in ${\cal C}^\ast$ as the candidate functions and follow the forward stepwise algorithm and then the backward stepwise algorithm as in Section \ref{sec2} to adaptively select the basis functions denoted by $\widehat h_{j}$, $j=1,\cdots,\widehat m$. By (\ref{DR}), we readily have that $
G(x)=G_0\left({\mathbf B}^{^\intercal}x\right)=G_0\left(x_\circ\right)$.	Since $x_\ast\rightarrow x_\circ$ by Theorem \ref{th:2}, instead of estimating $G$, we next estimate the nonparametric link function $G_0$ using the drMARS selected basis functions.

Let
\[\widehat{\mathbf H}(\cdot)=\left[1,\widehat h_1(\cdot),\cdots, \widehat h_{\widehat m}(\cdot)\right]^{^\intercal},\ \ \widehat{\mathbb H}_\ast=\left[\widehat{\mathbf H}(X_1^\ast),\cdots,\widehat{\mathbf H}(X_n^\ast)\right]^{^\intercal}.\]
We estimate the parameters in the basis expansion via least squares, i.e.,
\begin{equation}\label{drMARS-est-1}
	\widehat{\boldsymbol\gamma}=\left(\widehat\gamma_0,\widehat\gamma_1,\cdots,\widehat\gamma_{\widehat m}\right)^{^\intercal}=\left(\widehat{\mathbb H}_{\ast}^{^\intercal}\widehat{\mathbb H}_\ast\right)^{-1}\widehat{\mathbb H}_{\ast}^{^\intercal}{\mathbf Y},
\end{equation}
and then obtain the drMARS estimate:
\begin{equation}\label{drMARS-est-2}
	\widehat{G}_0(x_\ast)=\widehat\gamma_0+\sum_{j=1}^{\widehat m} \widehat\gamma_j\widehat h_j(x_\ast)=\widehat{\mathbf H}(x_\ast)^{^\intercal}\left(\widehat{\mathbb H}_{\ast}^{^\intercal}\widehat{\mathbb H}_\ast\right)^{-1}\widehat{\mathbb H}_{\ast}^{^\intercal}{\mathbf Y}.
\end{equation}
The main difference between drMARS and the conventional MARS in \cite{F91} is that the former incorporates the linear combinations of covariates determined by the SDR projection in the estimation algorithm. Hence drMARS is expected to work better when the underlying model contains the multiple-index structure (\ref{DR}). In particular, if a high-order interaction of covariates can be written as the multiple-index form, drMARS can significantly reduce the number of basis functions in the model approximation, and subsequently improve the estimation efficiency; see the simulation studies in Section \ref{sec4}.

Similar to ${\mathbf H}(\cdot)$ defined in Section \ref{sec2}, we let $\overline{\mathbf H}(\cdot)$ be a vector containing all the basis functions which are either from ${\mathcal C}^\ast$ or the tensor products of the marginal basis functions, i.e., 
\[\overline{\mathbf H}(\cdot)=\left[\widehat{\mathbf H}(\cdot)^{^\intercal}, \widehat {\mathbf H}_{-}(\cdot)^{^\intercal}\right]^{^\intercal},\]
where $\widehat{\mathbf H}_{-}(\cdot)$ is a vector of basis functions not selected by drMARS. Let $\overline{m}_{\overline{H}}$ be the dimension of $\overline{\mathbf H}(\cdot)$. Note that $\overline{\mathbf H}(\cdot)$ is a vector of deterministic functions, facilitating the asymptotic derivation of the drMARS estimation.

We need the following technical conditions to derive the convergence theory of the drMARS estimation.

\begin{assumption}\label{ass:2}
	
	{\em (i)\ The density function of $X_i^\circ$ exists, and is bounded away from zero and infinity on a compact set. The link function $G_0$ is continuous and differentiable.
	
	(ii)\ The matrix $\overline{\boldsymbol\Omega}:={\sf E}\left[\overline{\mathbf H}(X_i^\circ) \overline{\mathbf H}(X_i^\circ)^{^\intercal}\right]$ is positive definite, and $\overline{m}_{\overline{H}}\sqrt{\log \overline{m}_{\overline{H}}}=o(n)$.

    (iii)\ The numbers of drMARS selected basis functions and MARS selected ones: $\widehat m$ and $\widetilde m$, satisfy that $\widehat m\left[\widetilde m^{1/2}n^{-1/2}+\widetilde\rho(\widetilde m)\right]=o_P(1)$.
 
	(iv)\ There exists $\widehat\rho(\cdot)$ satisfying $\widehat\rho(u)\rightarrow0$ as $u\rightarrow\infty$, such that 
 \[\left\vert G_0(x_\ast)-\widehat{\mathbf H}(x_\ast)^{^\intercal}{\boldsymbol\gamma}_\ast\right\vert=O_P\left(\widehat\rho(m_\circ)\right),\ \ {\boldsymbol\gamma}_\ast=\left(\widehat{\mathbb H}_{n}^{\ast{^\intercal}}\widehat{\mathbb H}_{n}^\ast\right)^{-1}\widehat{\mathbb H}_{n}^{\ast{^\intercal}}{\mathbf G},\]
conditional on $\widehat m=m_\circ$.}
 
\end{assumption}

\begin{remark}\label{re:4}
	
	Assumption \ref{ass:2}(i) extends Assumption \ref{ass:1}(ii) to the setting including projected covariates. As discussed in Remark \ref{re:1}, Assumption \ref{ass:2}(ii) ensures that the least squares estimate (\ref{drMARS-est-1}) is well defined. In fact, Lemma \ref{le:3} in the appendix shows that $\frac{1}{n}\widehat{\mathbb H}_{\ast}^{^\intercal}\widehat{\mathbb H}_\ast$ is positive definite w.p.a.1, indicating that its inverse matrix exists. Let $\widehat{\cal S}$ be the estimation space by including the linear spline functions and their tensor products selected by drMARS. We may show that Assumption \ref{ass:2}(ii) can be replaced by the empirical identifiability condition on $\widehat{\cal S}$. Assumption \ref{ass:2}(iii), combined with the convergence property in Theorem \ref{th:2}, is crucial to ensure the consistency property when we replace $\widetilde{\mathbf B}$ by ${\mathbf B}$ in drMARS. 

    We next discuss the high-level condition in Assumption \ref{ass:2}(iv) on the drMARS estimation bias. Letting
    \[
    \overline{\mathbb H}_\circ=\left[\overline{\mathbf H}(X_1^\circ),\cdots, \overline{\mathbf H}(X_n^\circ)\right]^{^\intercal},\quad \widehat{\mathbb H}_\circ=\left[\widehat{\mathbf H}(X_1^\circ),\cdots, \widehat{\mathbf H}(X_n^\circ)\right]^{^\intercal},
    \]
    we define 
    \[
    {\boldsymbol\gamma}_\dagger=(\overline{\mathbb H}_\circ^{^\intercal}\overline{\mathbb H}_\circ)^{-1}\overline{\mathbb H}_\circ^{^\intercal}{\mathbf G},\quad {\boldsymbol\gamma}_\circ=(\widehat{\mathbb H}_\circ^{^\intercal}\widehat{\mathbb H}_\circ)^{-1}\widehat{\mathbb H}_\circ^{^\intercal}{\mathbf G}. 
    \]
    Similar to the bias decomposition (\ref{bias-1}) in Remark \ref{re:1}, we have
     \begin{eqnarray}
        G_0(x_\ast)-\widehat{\mathbf H}(x_\ast)^{^\intercal}{\boldsymbol\gamma}_\ast&=&\left[G_0(x_\circ)-\widehat{\mathbf H}(x_\circ)^{^\intercal}{\boldsymbol\gamma}_\circ\right]+\left[G_0(x_\ast)-\widehat{\mathbf H}(x_\ast)^{^\intercal}{\boldsymbol\gamma}_\ast-G_0(x_\circ)+\widehat{\mathbf H}(x_\circ)^{^\intercal}{\boldsymbol\gamma}_\circ\right]\notag\\
        &=&\left[G_0(x_\circ)-\overline{\mathbf H}(x_\circ)^{^\intercal}{\boldsymbol\gamma}_\dagger\right]+\left[\overline{\mathbf H}(x_\circ)^{^\intercal}{\boldsymbol\gamma}_\dagger-\widehat{\mathbf H}(x_\circ)^{^\intercal}{\boldsymbol\gamma}_\circ\right]\notag\\
        &&\left[G_0(x_\ast)-\widehat{\mathbf H}(x_\ast)^{^\intercal}{\boldsymbol\gamma}_\ast-G_0(x_\circ)+\widehat{\mathbf H}(x_\circ)^{^\intercal}{\boldsymbol\gamma}_\circ\right]\notag\\
        &=:&\widehat{\rm bias}_1\left(x\right)+\widehat{\rm bias}_2\left(x\right)+\widehat{\rm bias}_3\left(x\right).\label{bias-2}
 \end{eqnarray}
 Hence the high-level bias order in Assumption \ref{ass:2}(iv) combines the three bias terms in the decomposition (\ref{bias-2}). The first term $\widehat{\rm bias}_1\left(x\right)$ is caused by the approximation error of $G_0(\cdot)$ by its projection onto $\overline{\cal S}$, an estimation space containing the linear spline functions and their tensor products as in $\overline{\mathbf H}(\cdot)$. As discussed in Remark \ref{re:1}, under some smoothness conditions on $G_0$ , we may show that $\widehat{\rm bias}_1\left(x\right)$ is of order $\overline m_{\overline H}^{-q/d}$, where $q$ is a positive number relevant to the smoothness level of $G_0(\cdot)$. The second term $\widehat{\rm bias}_2\left(x\right)$ is induced by the projection onto the drMARS-selected estimation space $\widehat{\cal S}$ rather than $\overline{\cal S}$. \cite{L13} discusses the order of $\widehat{\rm bias}_2\left(x\right)$ under some extra restrictions. As discussed in Remark \ref{re:1}, if $\widehat m$ is of the same order as $\overline m_{\overline H}$, we conjecture that $\widehat{\rm bias}_2\left(x\right)$ would have the same approximation order as $\widehat{\rm bias}_1\left(x\right)$. Finally, $\widehat{\rm bias}_3\left(x\right)$ is the extra bias due to the replacement of ${\mathbf B}$ by $\widetilde{\mathbf B}$ in the drMARS algorithm. 
	
\end{remark}	


The following theorem gives the point-wise convergence rate for $\widehat{G}_0(x_\ast)$ defined in (\ref{drMARS-est-2}).

\smallskip

\begin{theorem}\label{th:3}
	
	Suppose that Assumptions \ref{ass:1} and \ref{ass:2} are satisfied. The drMARS estimate $\widehat{G}_0(x_\ast)$ has the following convergence result:
	\begin{equation}\label{th3}
	\widehat{G}_0(x_\ast)- G_0(x_\ast)=O_P\left( m_\circ^{1/2}/n^{1/2}+\widehat\rho(m_\circ)\right),
	\end{equation}
	conditional on $\widehat m=m_\circ$.
	
\end{theorem}


\begin{remark}\label{re:5}
	
	The convergence rate obtained in Theorem \ref{th:3} is comparable to those derived by \cite{H03} and \cite{L13} for the polynomial spline regression estimation. Assume that the nonparametric link function is sufficiently smooth, say $G_0(\cdot)$ is $q$-smooth \citep[e.g.,][]{H03}, $\widehat m\propto \overline m_{\overline H}$, and the convergence of $\widetilde{\mathbf B}$ is sufficiently fast, say $\widetilde{\mathbf B}$ is root-$n$ convergent \citep[e.g.,][]{X08}. Following the discussion in Remark \ref{re:4}, we conjecture that $\widehat\rho(m_\circ)$ is dominated by $\widehat{\rm bias}_1\left(x\right)+\widehat{\rm bias}_2\left(x\right)$, which is  upper bounded by $m_\circ^{-q/d}$ conditional on $\widehat m=m_\circ$. Consequently the point-wise convergence rate of drMARS becomes $O_P(m_\circ^{1/2}/n^{1/2}+m_\circ^{-q/d})$, indicating that the optimal order of $\widehat m=m_\circ$ is $n^{d/(d+2q)}$ and the optimal convergence rate is expected to be $O_P(n^{-q/(d+2q)})$ \citep[e.g.,][]{S82}. In contrast, as discussed in \cite{L13}, if $G(\cdot)$ is $q$-smooth, the conventional MARS estimation (without SDR rotation) has the bias order $\widehat m_\diamond^{-q/d}$, where $\widehat m_\diamond$ is the number of the MARS selected basis functions. If $\widehat m_\diamond$ has the optimal order $n^{p/(p+2q)}$, the point-wise convergence rate of the conventional MARS is $O_P\left(n^{-q/(p+2q)}\right)$. As $d$ is typically smaller than $p$, it is sensible to expect that drMARS has faster convergence rate than the conventional MARS under the multiple-index model framework (\ref{DR}). This is confirmed by the numerical studies in Section \ref{sec4} for finite samples.
	
\end{remark}

In practice, we may further modify drMARS to obtain the nonparametric estimation that is robust to possible model misspecification, i.e., the multiple-index structural assumption (\ref{DR}) is violated. Let $\check{X}_i=\left(X_i^{^\intercal}, X_i^{\ast{^\intercal}}\right)^{^\intercal}$ be a vector combining both the original and projected covariates. Consider a sample $(Y_1,\check X_1),\cdots,(Y_n, \check X_n)$, use the linear spline functions in ${\cal C}\cup{\cal C}^\ast$ as the candidate functions and apply MARS to adaptively select the basis functions denoted by $\check h_{j}$, $j=1,\cdots, \check m$. Similarly to (\ref{drMARS-est-2}), the estimate of $G(x)$ is obtained by
\begin{equation}\label{drMARS-est-3}
	\check{G}(x)=\check\phi_0+\sum_{j=1}^{\check m} \check\phi_j\cdot\check h_j(\check x),
\end{equation}
where $\check\phi_0, \check\phi_1,\cdots, \check\phi_{\check m}$ are the least squares estimates and $\check x=\left(x^{^\intercal}, x_\ast^{^\intercal}\right)^{^\intercal}$ with $x_\ast=\widetilde{\mathbf B}^{^\intercal} x$. Furthermore, the nonparametric estimate can be recast into the following form:
\[
\check{G}(x)=\check\phi_0+\check{G}_\dagger(x_\ast)+\check{G}_\ddagger(x),
\]
where $\check{G}_\dagger(x_\ast)$ is defined by summing over the terms in (\ref{drMARS-est-3}) whose basis functions involve only the projected covariates whereas $\check{G}_\ddagger(x)$ is defined by summing over the terms whose basis functions involve the original covariates. When the multiple-index model assumption is valid, it is expected that most of the basis functions involved in defining $\check{G}_\ddagger(x)$ would be screened out in the adaptive selection process, and consequently $\check{G}(x)$ is approximated by $\check\phi_0+\check{G}_\dagger(x_\ast)$, which is expected to be close to $\widehat{G}_0(x_\ast)$ defined in (\ref{drMARS-est-2}).

\section{Simulation studies}\label{sec4}

In this section, we use simulated data to showcase the performance of the proposed dimension reduction and drMARS methods in two aspects: estimation of the (central mean) SDR space and estimation of the regression function. For the SDR space estimation, we compare drMARS with principal Hessian directions \citep[pHd,][]{cook04}, conditional variance estimator \citep[CVE,][]{fertl22}, gradient-based kernel dimension reductiong \citep[gKDR,][]{FL14} and minimum average variance estimation \citep[MAVE,][]{XTLZ02}. The accuracy of an estimate $\mathbf{\widetilde B}$ is evaluated by 
\[
D(\mathbf{\widetilde B},\mathbf B) =  \left\Vert (\mathbf I - \mathbf B (\mathbf B^{^\intercal}\mathbf B)^{-1} \mathbf B^{^\intercal}) \mathbf{\widetilde B} \right\Vert_{\rm F} / \sqrt{d},
\]
where $d$ is the effective dimension which is assumed to be known. Selection of this dimension is evaluated separately. The smaller $D(\mathbf{\widetilde B},\mathbf B)$ is, the better the SDR space estimate is. For the regression function estimation, we compare drMARS with two popular methods: the support vector machine \citep[SVM,][]{cortes95} and random forest \citep[RF,][]{breiman01}. We also compare the function estimation using the SDR directions obtained by pHd, CVE, gKDR, MAVE and our drMARS, respectively, and call them SDR-MARS in general. For any estimate of the regression function $G(x) ={\sf E}(Y|X=x) $, say $\widehat G(x) $, we define the mean squared error (MSE):
\begin{equation*}
	{\sf MSE}(G)= \frac{1}{m}\sum_{i=1}^m \left[\widehat G(Z_i) - G(Z_i)\right]^2,
\end{equation*}
to evaluate the estimation accuracy, where, $\{X_1, \cdots, X_n\}$ is the in-sample used to estimate the regression function $G(\cdot)$ and $\{Z_1, \cdots, Z_m\}$ is the out-of-sample used to compute the MSE. Both follow the same distribution.

All methods are implemented with R. Specifically, package \texttt{dr} \citep{Weisberg02} for pHd, \texttt{cve} function in package \texttt{CVarE}  for CVE, package \texttt{MAVE} for MAVE, package \texttt{earth} \citep{earth23} for MARS, \texttt{svm} function in package \texttt{e1071} \citep{dimitriadou08} for SVM, package \texttt{randomForest} \citep{liaw15} for RF are used in our numerical studies. The source codes for gKDR and drMARS as well as all the relevant files can be downloaded from \url{https://github.com/liuyu-star/drMARS}. For all the R functions, their default values of tuning parameters are used. In addition, as the random rotation is a commonly-used ensemble method \citep[e.g.,][]{BlaserFryzlewicz2016, cannings2017random, bagnall2020rotation}, we also include it in our comparison, denoted by RAND. In our setting, for each random rotation matrix ${\mathbf B}$, RAND applies MARS to $ ({\mathbf B}^{^\intercal} X_i, Y_i), i = 1, \cdots, n $ to train the model and then predict the testing data.

The data is generated by the following nonlinear regression model:
\[Y_i=G(X_i)+ \varepsilon_i,\]	
where $X_i=\left(X_{i1},\cdots,X_{ip}\right)^{^\intercal}\stackrel{i.i.d.}\sim \mathsf{U_p}(-1,1)$ or $\mathsf{N_p}\left({\mathbf 0}_p, {\boldsymbol\Sigma}_X\right)$ with ${\boldsymbol\Sigma}_X=\left(0.6^{|i-j|}\right)_{p\times p}$, and $\varepsilon_i  \stackrel{i.i.d.}\sim \mathsf{N}(0,0.5^2)$. The specifications of $G(\cdot)$ are as follows,
\begin{eqnarray*}
	({\rm M1})&&\  G(x)=0.5(x_{1}+x_{2})+2.5\exp(-2(x_{1}+ x_{2}+x_{3})^2),\\
	({\rm M2})&&\ G(x)=\frac{1}{30} \exp(4x_{1})+\frac{4}{3+3\exp(-20(x_{2}-0.5))}+ \frac{3x_{3}+2x_{4}+x_{5}}{3},\\
	({\rm M3})&&\ G(x)=0.6\sin(\pi x_{1}x_{2})+1.2(x_{3}-0.5)^2+0.6x_{4}+0.3x_{5},  \\
	({\rm M4})&&\  G(x)= 5x_{1} x_{2}x_{3},\\
	({\rm M5})&&\ G(x)=4(x_{1}-x_{2}+x_{3}) \sin(0.5\pi(x_{1}+x_{2})),\\
	({\rm M6})&&\ G(x)=x_{1} (x_{1}+x_{2}+1),\\
	({\rm M7})&&\ G(x)=\frac{x_{1}} {0.5+(x_{2}+1.5)^2}.
\end{eqnarray*}
In terms of SDR, the effective dimensions for M1, $\cdots$, M7 are 2, 3, 4, 3, 2, 2 and 2, respectively. For example, the SDR space of M3 is spanned by $\beta_1=\left(1, {\mathbf 0}_{p-1}^{^\intercal}\right)^{^\intercal}, \beta_2=\left(0, 1, {\mathbf 0}_{p-2}^{^\intercal}\right)^{^\intercal}, \beta_3=\left(0, 0, 1, {\mathbf 0}_{p-3}^{^\intercal}\right)^{^\intercal} \text{ and } \beta_4=\left(0, 0, 0, 2/\sqrt{5}, 1/\sqrt{5}, {\mathbf 0}_{p-5}^{^\intercal}\right)^{^\intercal}$. {The dimension $p$ is  $50$ or $100$}, and the sample size $n$ is $200$ or $500$.

\begin{table}[h!]
	\centering
	\caption{{Average $D(\mathbf{\widetilde B},\mathbf B)$ for estimation of the SDR space over $100$ replications: the smaller the value, the better the method.}}\label{DBB}%
	\begin{minipage}{\textwidth}
		\setlength{\tabcolsep}{1.0mm}{
			\begin{tabular*}{\textwidth}{@{\extracolsep{\fill}}lcccccccccccc@{\extracolsep{\fill}}}
				\toprule%
				&       &         & \multicolumn{5}{@{}c@{}}{$X\sim \mathsf{U_p}(-1,1)$} & \multicolumn{5}{@{}c@{}}{$X\sim \mathsf{N_p}({\mathbf 0}_p, {\boldsymbol\Sigma}_X)$} \\
				\cmidrule(lr){4-8}\cmidrule(lr){9-13}%
				Model  & $p$     & $n$& pHd & CVE &gKDR & MAVE& drMARS &pHd & CVE &gKDR & MAVE& drMARS\\
			\midrule
			\multirow{4}[4]{*}{M1} & \multirow{2}[2]{*}{50} & 200   & 0.89  & 0.75  & 0.83  & 0.74  & 0.53  & 0.98  & 0.76  & 0.83  & 0.87  & 0.80  \\
			&       & 500   & 0.74  & 0.66  & 0.69  & 0.60  & 0.47  & 0.96  & 0.71  & 0.75  & 0.79  & 0.71  \\
			\cdashline{2-13}          & \multirow{2}[2]{*}{100} & 200   & 0.98  & 0.94  & 0.91  & 0.75  & 0.58  & 0.99  & 0.83  & 0.87  & 0.87  & 0.85  \\
			&       & 500   & 0.86  & 0.71  & 0.81  & 0.69  & 0.44  & 0.99  & 0.74  & 0.82  & 0.87  & 0.73  \\
			\hline
			\multirow{4}[4]{*}{M2} & \multirow{2}[2]{*}{50} & 200   & 0.96  & 0.84  & 0.82  & 0.81  & 0.39  & 0.96  & 0.85  & 0.76  & 0.90  & 0.81  \\
			&       & 500   & 0.94  & 0.81  & 0.77  & 0.72  & 0.34  & 0.95  & 0.85  & 0.66  & 0.91  & 0.79  \\
			\cdashline{2-13}          & \multirow{2}[2]{*}{100} & 200   & 0.98  & 0.89  & 0.86  & 0.81  & 0.38  & 0.98  & 0.91  & 0.83  & 0.89  & 0.82  \\
			&       & 500   & 0.97  & 0.84  & 0.82  & 0.82  & 0.31  & 0.98  & 0.90  & 0.75  & 0.94  & 0.83  \\
			\hline
			\multirow{4}[4]{*}{M3} & \multirow{2}[2]{*}{50} & 200   & 0.94  & 0.88  & 0.85  & 0.84  & 0.72  & 0.93  & 0.90  & 0.83  & 0.82  & 0.38  \\
			&       & 500   & 0.89  & 0.84  & 0.78  & 0.68  & 0.66  & 0.89  & 0.86  & 0.77  & 0.76  & 0.20  \\
			\cdashline{2-13}          & \multirow{2}[2]{*}{100} & 200   & 0.98  & 0.94  & 0.90  & 0.86  & 0.77  & 0.97  & 0.95  & 0.93  & 0.81  & 0.46  \\
			&       & 500   & 0.96  & 0.89  & 0.86  & 0.84  & 0.70  & 0.94  & 0.90  & 0.86  & 0.82  & 0.25  \\
			\hline
			\multirow{4}[4]{*}{M4} & \multirow{2}[2]{*}{50} & 200   & 0.94  & 0.92  & 0.96  & 0.95  & 0.78  & 0.96  & 0.85  & 0.87  & 0.86  & 0.21  \\
			&       & 500   & 0.93  & 0.81  & 0.92  & 0.95  & 0.51  & 0.95  & 0.83  & 0.78  & 0.82  & 0.04  \\
			\cdashline{2-13}          & \multirow{2}[2]{*}{100} & 200   & 0.98  & 0.97  & 0.98  & 0.95  & 0.85  & 0.98  & 0.89  & 0.96  & 0.86  & 0.33  \\
			&       & 500   & 0.96  & 0.96  & 0.98  & 0.98  & 0.65  & 0.98  & 0.84  & 0.89  & 0.88  & 0.08  \\
			\hline
			\multirow{4}[4]{*}{M5} & \multirow{2}[2]{*}{50} & 200   & 0.81  & 0.78  & 0.97  & 0.42  & 0.31  & 0.93  & 0.93  & 0.98  & 0.96  & 0.79  \\
			&       & 500   & 0.50  & 0.21  & 0.77  & 0.11  & 0.49  & 0.90  & 0.85  & 0.93  & 0.94  & 0.44  \\
			\cdashline{2-13}          & \multirow{2}[2]{*}{100} & 200   & 0.97  & 0.96  & 0.99  & 0.66  & 0.27  & 0.97  & 0.97  & 0.99  & 0.95  & 0.84  \\
			&       & 500   & 0.75  & 0.71  & 0.99  & 0.21  & 0.47  & 0.94  & 0.94  & 0.99  & 0.98  & 0.54  \\
			\hline
			\multirow{4}[4]{*}{M6} & \multirow{2}[2]{*}{50} & 200   & 0.95  & 0.76  & 0.73  & 0.71  & 0.57  & 0.92  & 0.74  & 0.92  & 0.69  & 0.28  \\
			&       & 500   & 0.83  & 0.61  & 0.56  & 0.59  & 0.56  & 0.84  & 0.71  & 0.69  & 0.61  & 0.08  \\
			\cdashline{2-13}          & \multirow{2}[2]{*}{100} & 200   & 0.98  & 0.87  & 0.83  & 0.71  & 0.57  & 0.98  & 0.78  & 0.96  & 0.67  & 0.40  \\
			&       & 500   & 0.95  & 0.73  & 0.68  & 0.70  & 0.57  & 0.92  & 0.73  & 0.91  & 0.70  & 0.13  \\
			\hline
			\multirow{4}[4]{*}{M7} & \multirow{2}[2]{*}{50} & 200   & 0.96  & 0.88  & 0.86  & 0.88  & 0.76  & 0.97  & 0.80  & 0.86  & 0.82  & 0.47  \\
			&       & 500   & 0.91  & 0.77  & 0.76  & 0.81  & 0.71  & 0.93  & 0.75  & 0.73  & 0.74  & 0.22  \\
			\cdashline{2-13}          & \multirow{2}[2]{*}{100} & 200   & 0.99  & 0.94  & 0.92  & 0.90  & 0.76  & 0.99  & 0.86  & 0.90  & 0.81  & 0.55  \\
			&       & 500   & 0.97  & 0.87  & 0.84  & 0.89  & 0.66  & 0.98  & 0.78  & 0.83  & 0.84  & 0.26  \\
			\bottomrule
			\end{tabular*}}
		\end{minipage}
\end{table}
For the estimation of the SDR space, the simulation results based on 100 replications are shown in Tables \ref{DBB} and \ref{dSDR}. For the seven models, Table \ref{DBB} shows that the estimation errors of drMARS are smaller than those of pHd, CVE, gKDR and MAVE, indicating that drMARS has significant improvement over the competing methods in estimating the SDR space. Moreover, in most cases the relative estimation error reduction of drMARS over the others improves as the dimension $p $ increases. For example, for M2 with $X$ following the uniform distribution and $n = 500$, the relative estimation error reduction (drMARS over MAVE) is $(0.72-0.34)/0.72=0.5278$ when $p = 50$, and it increases to $(0.82-0.31)/0.82=0.6220$ when $p = 100$.
\begin{table}[h!]
	\centering
	\caption{The proportion of selecting the true dimension of the SDR space $\rho(\widehat{d}=d_0)$ and the computing time for estimating the SDR space (with the true dimension) over $100$ replications when $X\sim \mathsf{N_p}({\mathbf 0}_p, {\boldsymbol\Sigma}_X)$.}\label{dSDR}%
	\begin{minipage}{\textwidth}
		\setlength{\tabcolsep}{1.0mm}{
			\begin{tabular*}{\textwidth}{@{\extracolsep{\fill}}lcccccccccccc@{\extracolsep{\fill}}}
				\toprule%
				&       &         & \multicolumn{5}{@{}c@{}}{$\rho(\widehat{d}=d_0)$} & \multicolumn{5}{@{}c@{}}{Computational time (in seconds)} \\
				\cmidrule(lr){4-8}\cmidrule(lr){9-13}%
				Model  & $p$     & $n$& pHd & CVE &gKDR & MAVE& drMARS &pHd & CVE &gKDR & MAVE& drMARS\\
				\hline
				\multirow{4}[4]{*}{M1} & \multirow{2}[2]{*}{50} & 200   & 0.00  & 0.14  & 0.21  & 0.13  & 0.31  & 0.01  & 5.79  & 0.33  & 2.08  & 1.57  \\
				&       & 500   & 0.00  & 0.00  & 0.17  & 0.09  & 0.25  & 0.01  & 27.56  & 1.55  & 7.42  & 3.38  \\
				\cdashline{2-13}          & \multirow{2}[2]{*}{100} & 200   & 0.00  & 0.31  & 0.19  & 0.18  & 0.19  & 0.02  & 17.70  & 0.70  & 2.11  & 2.78  \\
				&       & 500   & 0.00  & 0.01  & 0.11  & 0.00  & 0.20  & 0.04  & 55.85  & 4.85  & 18.70  & 6.43  \\
				\hline
				\multirow{4}[4]{*}{M2} & \multirow{2}[2]{*}{50} & 200   & 0.15  & 0.08  & 0.10  & 0.12  & 0.17  & 0.01  & 8.42  & 0.28  & 1.87  & 0.82  \\
				&       & 500   & 0.13  & 0.10  & 0.08  & 0.14  & 0.24  & 0.01  & 47.90  & 1.84  & 6.71  & 2.22  \\
				\cdashline{2-13}          & \multirow{2}[2]{*}{100} & 200   & 0.17  & 0.10  & 0.04  & 0.13  & 0.21  & 0.02  & 22.04  & 0.79  & 1.89  & 1.24  \\
				&       & 500   & 0.13  & 0.14  & 0.03  & 0.17  & 0.20  & 0.03  & 76.34  & 5.06  & 16.95  & 3.17  \\
				\hline
				\multirow{4}[4]{*}{M3} & \multirow{2}[2]{*}{50} & 200   & 0.21  & 0.00  & 0.35  & 0.05  & 0.36  & 0.01  & 13.05  & 0.26  & 1.90  & 1.50  \\
				&       & 500   & 0.24  & 0.00  & 0.27  & 0.06  & 0.30  & 0.01  & 66.94  & 1.53  & 6.77  & 4.01  \\
				\cdashline{2-13}          & \multirow{2}[2]{*}{100} & 200   & 0.30  & 0.00  & 0.29  & 0.01  & 0.27  & 0.02  & 41.58  & 0.74  & 1.95  & 2.82  \\
				&       & 500   & 0.15  & 0.00  & 0.23  & 0.10  & 0.34  & 0.03  & 134.85  & 4.65  & 17.25  & 7.32  \\
				\hline
				\multirow{4}[4]{*}{M4} & \multirow{2}[2]{*}{50} & 200   & 0.13  & 0.01  & 0.23  & 0.14  & 0.33  & 0.01  & 9.53  & 0.29  & 1.91  & 1.13  \\
				&       & 500   & 0.15  & 0.00  & 0.20  & 0.04  & 0.36  & 0.01  & 49.21  & 1.83  & 6.76  & 3.09  \\
				\cdashline{2-13}          & \multirow{2}[2]{*}{100} & 200   & 0.17  & 0.00  & 0.10  & 0.12  & 0.35  & 0.02  & 28.93  & 0.82  & 1.93  & 1.96  \\
				&       & 500   & 0.16  & 0.00  & 0.18  & 0.11  & 0.43  & 0.04  & 93.66  & 5.18  & 17.24  & 5.41  \\
				\hline
				\multirow{4}[4]{*}{M5} & \multirow{2}[2]{*}{50} & 200   & 0.04  & 0.16  & 0.17  & 0.20  & 0.26  & 0.01  & 6.43  & 0.25  & 2.08  & 1.35  \\
				&       & 500   & 0.00  & 0.14  & 0.20  & 0.15  & 0.40  & 0.01  & 33.98  & 1.44  & 7.38  & 4.33  \\
				\cdashline{2-13}          & \multirow{2}[2]{*}{100} & 200   & 0.04  & 0.14  & 0.01  & 0.15  & 0.30  & 0.02  & 19.38  & 0.70  & 2.12  & 2.61  \\
				&       & 500   & 0.00  & 0.14  & 0.15  & 0.04  & 0.29  & 0.03  & 63.76  & 4.44  & 18.68  & 6.75  \\
				\hline
				\multirow{4}[4]{*}{M6} & \multirow{2}[2]{*}{50} & 200   & 0.23  & 0.02  & 0.07  & 0.33  & 0.44  & 0.01  & 5.37  & 0.26  & 2.07  & 1.01  \\
				&       & 500   & 0.14  & 0.01  & 0.08  & 0.50  & 0.46  & 0.01  & 25.99  & 1.53  & 7.39  & 1.42  \\
				\cdashline{2-13}          & \multirow{2}[2]{*}{100} & 200   & 0.13  & 0.01  & 0.16  & 0.38  & 0.33  & 0.02  & 17.86  & 0.71  & 2.11  & 1.93  \\
				&       & 500   & 0.21  & 0.05  & 0.01  & 0.25  & 0.47  & 0.04  & 49.32  & 4.70  & 18.63  & 2.78  \\
				\hline
				\multirow{4}[4]{*}{M7} & \multirow{2}[2]{*}{50} & 200   & 0.03  & 0.18  & 0.19  & 0.21  & 0.35  & 0.01  & 5.43  & 0.25  & 2.08  & 1.38  \\
				&       & 500   & 0.03  & 0.21  & 0.30  & 0.25  & 0.50  & 0.02  & 27.11  & 1.43  & 7.39  & 3.48  \\
				\cdashline{2-13}          & \multirow{2}[2]{*}{100} & 200   & 0.09  & 0.18  & 0.06  & 0.16  & 0.39  & 0.02  & 17.09  & 0.70  & 2.11  & 2.76  \\
				&       & 500   & 0.05  & 0.23  & 0.02  & 0.07  & 0.53  & 0.04  & 51.15  & 4.28  & 18.71  & 6.60  \\
				\bottomrule
		\end{tabular*}}
	\end{minipage}
\end{table}

As mentioned in Section \ref{sec2}, we select the dimension of SDR space by the 10-fold CV criterion, using a similar idea as in \cite{XTLZ02}. The data sample is randomly divided into $10$ equal subsamples ${\cal I}_k$ with size $\lfloor0.1n\rfloor$, i.e.,  $\{1, \cdots, n\}=\cup_{k=1}^{10}{\cal I}_k$. The true dimension (denoted by $d_0$) is estimated as follows
$$
\widehat{d}=\mathop{\arg\max}\limits _{1\leq d\leq \overline d} {\sf CV}(d),\quad {\sf CV}(d)=\frac{1}{10}\sum_{k=1}^{10} R^2(d, {\cal I}_k),
$$
where $\overline{d}$ is set as $5$ in the simulation,
$$
R^2(d, {\cal I}_k)=1- \frac{\sum_{i \in {\cal I}_k}\left(Y_i-\widehat{G}_0^{-{\cal I}_k}(\widehat{\mathbf{B}}_d^{^\intercal} \mathbf{X}_{i})\right)^2}
 {\sum_{i \in {\cal I}_k}\left(Y_i-\overline{Y}^{-{\cal I}_k}\right)^2},
$$
 $\widehat{\mathbf{B}}_d$ is computed from the whole data set by setting the dimension of SDR space as $d$, $\widehat{G}_0^{-{\cal I}_k}(\cdot)$ is computed from the data $\{(Y_i, \widehat{\mathbf{B}}_d^{^\intercal} \mathbf{X}_i) : i \notin {\cal I}_k \}$ using MARS in the R package \texttt{earth}, and $\overline{Y}^{-{\cal I}_k}$ is the average value of the response $\{ Y_i : i \notin {\cal I}_k\}$. The frequencies of correctly selecting the correct dimension and the computational time for estimating the SDR space over $100$ replications are reported in Table \ref{dSDR}, where only the results for $X\sim \mathsf{N_p}({\mathbf 0}_p, {\boldsymbol\Sigma}_X)$ are reported as the performance is similar for uniformly distributed covariates. For most cases, the dimension estimates based on drMARS are the most accurate one with $\rho(\widehat d=d_0)$ larger than the other methods. Regarding the computing time, pHd is the least time consuming and CVE is the most time consuming, whereas drMARS is in the middle. In summary, drMARS can substantially improve dimension estimation accuracy with reasonable (and acceptable) computational time.

\begin{table}[h]
	\centering
		\caption{Average {\sf MSE}($G$) for the regression function estimation over $100$ replications when $X\sim \mathsf{U_p}(-1,1)$.}\label{MSE}%
	\begin{minipage}{\textwidth}
		\setlength{\tabcolsep}{0.8mm}{
			\begin{tabular*}{\textwidth}{@{\extracolsep{\fill}}lrrrrrrrrrrr@{\extracolsep{\fill}}}
				\toprule%
				&       &         & \multicolumn{3}{@{}c@{}}{Original} & \multicolumn{6}{@{}c@{}}{SDR-MARS} \\
				\cmidrule(lr){4-6}\cmidrule(lr){7-12}%
				Model  & $p$     & $n$   & SVM   & RF    & MARS  & RAND  & pHd & CVE  & gKDR  & MAVE  & drMARS \\
				\hline
				\multirow{4}[4]{*}{M1} & \multirow{2}[2]{*}{50} & 200   & 0.95  & 0.80  & 0.99  & 0.92  & 1.03  & 0.49  & 1.02  & 0.54  & 0.35  \\
				&       & 500   & 0.87  & 0.61  & 0.52  & 0.82  & 0.46  & 0.15  & 0.51  & 0.13  & 0.09  \\
				\cdashline{2-12}          & \multirow{2}[2]{*}{100} & 200   & 0.98  & 0.83  & 1.23  & 0.98  & 1.65  & 1.42  & 1.45  & 0.63  & 0.40  \\
				&       & 500   & 0.95  & 0.66  & 0.64  & 0.92  & 0.74  & 0.26  & 0.66  & 0.32  & 0.10  \\
				\hline
				\multirow{4}[4]{*}{M2} & \multirow{2}[2]{*}{50} & 200   & 0.43  & 0.28  & 0.42  & 0.35  & 0.42  & 0.33  & 0.36  & 0.59  & 0.36  \\
				&       & 500   & 0.28  & 0.16  & 0.29  & 0.23  & 0.29  & 0.27  & 0.28  & 0.34  & 0.27  \\
				\cdashline{2-12}          & \multirow{2}[2]{*}{100} & 200   & 0.61  & 0.31  & 0.43  & 0.55  & 0.43  & 0.54  & 0.62  & 0.61  & 0.38  \\
				&       & 500   & 0.38  & 0.18  & 0.34  & 0.34  & 0.34  & 0.31  & 0.33  & 0.57  & 0.31  \\
				\hline
				\multirow{4}[4]{*}{M3} & \multirow{2}[2]{*}{50} & 200   & 0.49  & 0.29  & 0.63  & 0.43  & 0.64  & 0.44  & 0.44  & 0.65  & 0.36  \\
				&       & 500   & 0.35  & 0.22  & 0.42  & 0.30  & 0.42  & 0.28  & 0.32  & 0.32  & 0.27  \\
				\cdashline{2-12}          & \multirow{2}[2]{*}{100} & 200   & 0.63  & 0.31  & 0.72  & 0.59  & 0.74  & 0.67  & 0.74  & 0.71  & 0.42  \\
				&       & 500   & 0.45  & 0.24  & 0.55  & 0.42  & 0.55  & 0.39  & 0.40  & 0.64  & 0.32  \\
				\hline
				\multirow{4}[4]{*}{M4} & \multirow{2}[2]{*}{50} & 200   & 0.98  & 0.98  & 2.20  & 1.00  & 1.22  & 1.28  & 1.10  & 1.91  & 1.11  \\
				&       & 500   & 0.99  & 0.96  & 1.25  & 0.97  & 0.92  & 0.62  & 0.93  & 1.10  & 0.70  \\
				\cdashline{2-12}          & \multirow{2}[2]{*}{100} & 200   & 0.97  & 0.96  & 2.45  & 0.99  & 1.73  & 1.49  & 1.73  & 1.99  & 1.41  \\
				&       & 500   & 0.98  & 0.96  & 1.94  & 0.97  & 1.24  & 1.27  & 1.13  & 1.84  & 0.97  \\
				\hline
				\multirow{4}[4]{*}{M5} & \multirow{2}[2]{*}{50} & 200   & 6.57  & 4.18  & 1.09  & 6.57  & 1.09  & 1.09  & 1.09  & 1.08  & 0.87  \\
				&       & 500   & 6.30  & 2.83  & 0.24  & 6.20  & 0.24  & 0.24  & 0.24  & 0.26  & 0.25  \\
				\cdashline{2-12}          & \multirow{2}[2]{*}{100} & 200   & 6.60  & 4.53  & 1.56  & 6.68  & 1.56  & 1.63  & 1.56  & 1.54  & 0.95  \\
				&       & 500   & 6.61  & 3.03  & 0.27  & 6.49  & 0.27  & 0.27  & 0.27  & 0.27  & 0.28  \\
				\hline
				\multirow{4}[4]{*}{M6} & \multirow{2}[2]{*}{50} & 200   & 0.34  & 0.15  & 0.36  & 0.30  & 0.37  & 0.24  & 0.34  & 0.32  & 0.15  \\
				&       & 500   & 0.25  & 0.09  & 0.26  & 0.21  & 0.26  & 0.10  & 0.20  & 0.16  & 0.09  \\
				\cdashline{2-12}          & \multirow{2}[2]{*}{100} & 200   & 0.40  & 0.16  & 0.40  & 0.39  & 0.40  & 0.47  & 0.52  & 0.36  & 0.19  \\
				&       & 500   & 0.31  & 0.09  & 0.31  & 0.30  & 0.32  & 0.20  & 0.27  & 0.34  & 0.10  \\
				\hline
				\multirow{4}[4]{*}{M7} & \multirow{2}[2]{*}{50} & 200   & 0.08  & 0.05  & 0.35  & 0.08  & 0.23  & 0.18  & 0.13  & 0.31  & 0.14  \\
				&       & 500   & 0.06  & 0.03  & 0.23  & 0.06  & 0.17  & 0.10  & 0.08  & 0.18  & 0.12  \\
				\cdashline{2-12}          & \multirow{2}[2]{*}{100} & 200   & 0.09  & 0.05  & 0.39  & 0.09  & 0.35  & 0.24  & 0.24  & 0.32  & 0.17  \\
				&       & 500   & 0.07  & 0.04  & 0.29  & 0.07  & 0.23  & 0.17  & 0.11  & 0.29  & 0.12  \\
				\bottomrule
			\end{tabular*}}
		\end{minipage}
\end{table}
Table \ref{MSE} lists the MSEs for nonparametric regression function estimation, where only the results for $X\sim \mathsf{U_p}(-1,1)$ are reported as the performance for Gaussian covariates is similar. Generally, drMARS has smaller MSEs than the conventional MARS. For example, for model M1 with $p=50$ and $n=500$, the {\sf MSE}($G$) of MARS is $0.52$, the {\sf MSE}($G$) of the SDR-MARS with SDR estimated by pHd, CVE, gKDR and MAVE are smaller, and the {\sf MSE}($G$) of our drMARS is the smallest. A similar pattern can also be found for the other data generating processes. Note that the {\sf MSE}($G$) of MARS may be larger than that of SVM and RF for some of the data generating processes (such as M1 and M4). However, in most cases, our drMARS has smaller {\sf MSE}($G$) than (or comparable {\sf MSE}($G$) to) the SVM and RF methods. The simulation results in Table \ref{MSE} also show that RAND has poorer numerical performance than the other SDR-MARS methods.

\section{Real data analysis}\label{sec5}

In this section, we apply the proposed drMARS to the out-of-sample prediction of real data and statistical inference. Similarly to the simulation studies in Section \ref{sec4}, we consider the conventional MARS and SDR-MARS using various SDR estimation methods and compare their performance with other commonly-used nonparametric regression methods such as RF and SVM. We build the model using the training  set $\left\{(X_i^{\rm train}, Y^{\rm train}_i): i = 1, \cdots, n\right\} $, and make prediction for the testing set $\left\{(X_i^{\rm test}, Y^{\rm test}_i): i = 1, \cdots, m\right\} $.  The prediction performance is evaluated by the relative mean squared prediction error:
$$
{\sf rMSPE} =  \sum_{i=1}^m \left(\widehat Y^{\rm test}_i - Y^{\rm test}_i\right)^2/ \sum_{i=1}^m \left(\overline Y - Y^{\rm test}_i\right)^2,
$$
where $\widehat Y^{\rm test}_i$ is the fitted value of the response $Y_i^{\rm test}$ and $\overline Y = \frac{1}{n} \sum_{i=1}^n Y_i^{\rm train} $ is a naive prediction using the average of response observations in the learning set. In addition, as the use of logistic regression, our drMARS can be used for classification with two categories denoted as $0$ and $1$. Specifically, letting the prediction for the testing set be $\widehat y^{\rm test}_i,  i=1, 2, \cdots, m$,  the classification is $\widehat Y^{\rm test}_i= \mathbb I\left(\widehat y^{\rm test}_i>0.5\right)$, where $\mathbb I(\cdot)$ is the indicator function. The classification performance is measured by the misclassification rate (MCR) defined as
\begin{equation*}
{\sf MCR} =\frac{1}{m} \sum_{i=1}^m \mathbb I(\widehat Y^{\rm test}_i \neq Y^{\rm test}_i).
\end{equation*}

The following data sets are used to demonstrate the performance of prediction.
\begin{description}
	\item[\bf data.1]The data (\url{https://archive.ics.uci.edu/ml/datasets/concrete+compressive+strength}) is about the concrete compressive strength ($Y$) and its dependence on concrete's ingredients and age ($X$). It has $p=8$ predictors and $N=1,030$ observations. The square root transformation is made to the concrete compressive strength as the response.
	\item[\bf data.2] The data (\url{www.kaggle.com/harlfoxem/housesalesprediction}) contains house sale prices for King County in US including Seattle between May 2014 and May 2015. It contains $N=21,613$ house sale records. The interest is to predict the house sale prices ($Y$) based on $p=18$ variables ($X$). The logarithm transformation is made to the house sale prices.
	\item[\bf data.3]The data (\url{archive.ics.uci.edu/ml/datasets/Parkinsons+Telemonitoring}) is composed of a range of biomedical voice measurements from 42 people with early-stage Parkinson's disease recruited to a six-month trial of a telemonitoring device for remote symptom progression monitoring. Data on people's age,  gender, time interval from baseline recruitment date and 16 biomedical voice measures are the covariates ($X$ with $p=19$), and $N=5,875$ voice recording from these individuals are collected.  Our interest is to predict the motor scores ('motor\_UPDRS', $Y$) from the 19 covariates.
	\item[\bf data.4] The data  (\url{https://archive.ics.uci.edu/ml/datasets/Residential+Building+Data+Set}) contains construction cost, project variables, and economic variables corresponding to real estate single-family residential apartments in Tehran, Iran. It contains $N=372$ observations. The interest is to predict the construction cost ($Y$) using $p=102$ predictors ($X$) without considering the construction year. The logarithm transformation is made to the construction cost.

The following data sets are used to demonstrate the classification performance.

	\item[\bf data.5]The data (\url{www.kaggle.com/datasets/muratkokludataset/pistachio-dataset})  includes a total of $N=2,148$ images, 1,232 of Kirmizi type ($Y=0$) and 916 of Siirt type ($Y=1$). Each image contains 12 morphological features, 4 shape features and 12 color features ($p=28$). We are interested in the classification of the images based on the 28 features.
	\item[\bf data.6]The data (\url{https://archive.ics.uci.edu/ml/datasets/Hill-Valley}) contains $N=1,212$ records, each of which represents $p=100$ points on a two-dimensional graph. When plotted in order (from $1$ through $100$) as the $Y$ co-ordinate, the points will create either a Hill (a "bump" in the terrain, $Y=1$) or a Valley (a "dip" in the terrain, $Y=0$). Our interest is to discriminate whether a given record is a Hill or a valley by $100$ points on a two-dimensional graph.
	\item[\bf data.7] The data ({\footnotesize \url{www.kaggle.com/datasets/cnic92/200-financial-indicators-of-us-stocks-20142018}}) includes $N=986$ US stocks in year 2018, each of which contains $p=216$ financial indicators. These predictors are commonly found in the 10-K filings each publicly traded company releases yearly.  Each stock is classified into two classes: if the value of a stock increases during 2019 then $Y=1$; if the value of a stock decreases during 2019 then $Y=0$. The interest is to classify those stocks that are buy-worthy or not.
\end{description}

We randomly select $n=\min(1000, \lfloor N/3 \rfloor)$ or $n=\min(2000, \lfloor 2N/3 \rfloor)$ observations as the training set, and the remaining observations as the testing set, and repeat the random splitting $100$ times. The dimension of SDR space is selected using the 10-fold CV described in Section \ref{sec4}. The average {\sf rMSPEs} and {\sf MCRs} are reported in Table \ref{rMSPE-MCR} below.

\begin{table}[h!]
	\centering
	\caption{{Average {\sf rMSPE} or {\sf MCR} of the real data over $100$ replications} (in \%)}\label{rMSPE-MCR}
	\begin{minipage}{\textwidth}
		\setlength{\tabcolsep}{1.0mm}{
			\begin{tabular*}{\textwidth}{@{\extracolsep{\fill}}ccrrrrrrrrr@{\extracolsep{\fill}}}
				\toprule%
				Data&   Training    & \multicolumn{3}{@{}c@{}}{Original} & \multicolumn{6}{@{}c@{}}{SDR-MARS} \\
				\cmidrule(lr){3-5}\cmidrule(lr){6-11}%
				$(N, p)$   & size  & SVM   & RF    & MARS  & RAND  & pHd   & CVE  & gKDR  & MAVE  & drMARS \\
				\midrule
				data.1 & 343   & 21.30  & 17.86  & 15.85  & 21.42  & 15.85  & 15.85  & 15.85  & 15.88  & 12.69  \\
				(1030, 8) & 686   & 16.75  & 12.05  & 14.21  & 20.10  & 14.21  & 14.21  & 14.15  & 13.90  & 10.72  \\
				\cdashline{1-11}
				data.2 & 1000  & 22.55  & 16.06  & 15.50  & 35.14  & 15.50  & 15.50  & 15.50  & 15.46  & 14.62  \\
				(21613, 18) & 2000  & 19.57  & 14.35  & 13.89  & 33.37  & 13.89  & 13.89  & 13.86  & 13.67  & 13.09  \\
				\cdashline{1-11}
				data.3 & 1000  & 67.14  & 35.05  & 32.16  & 70.75  & 30.43  & 31.14  & 31.08  & 31.05  & 12.79  \\
				(5875, 19) & 2000  & 60.20  & 23.98  & 30.62  & 69.58  & 26.43  & 28.95  & 26.51  & 28.37  & 10.60  \\
				 \cdashline{1-11}
				data.4 & 124   & 10.02  & 8.74  & 5.85  & 28.86  & 12.82  & 5.83  & 5.86  & 6.01  & 4.75  \\
				(372, 102) & 248   & 6.73  & 6.41  & 4.21  & 26.75  & 7.93  & 4.17  & 4.22  & 4.34  & 3.68  \\
			   \hline
				data.5 & 716   & 8.74  & 11.22  & 9.30  & 11.23  & 9.30  & 9.30  & 9.30  & 8.27  & 9.07  \\
				(2148,28) & 1432  & 7.55  & 10.31  & 7.72  & 10.28  & 7.72  & 7.72  & 7.72  & 7.29  & 7.52  \\
				\cdashline{1-11}
				data.6 & 404   & 50.00  & 44.88  & 19.38  & 8.39  & 20.08  & 17.88  & 8.31  & 14.60  & 6.21  \\
				(1212, 100) & 808   & 50.07  & 40.88  & 19.57  & 5.68  & 16.90  & 17.03  & 6.24  & 17.08  & 3.16\\
				\cdashline{1-11}
				data.7 & 328   & 19.47  & 5.19  & 0.35  & 21.68  & 0.49  & 0.35  & 0.36  & 0.95  & 0.31  \\
				(986 ,216) & 657   & 13.94  & 0.94  & 0.12  & 21.46  & 0.13  & 0.12  & 0.12  & 0.35  & 0.11  \\
				\bottomrule
		\end{tabular*}}
	\end{minipage}
\end{table}

As can be seen from Table \ref{rMSPE-MCR}, for data.2 and data.4, the conventional MARS has smaller {\sf rMSPE} than SVM and RF. SDR-MARS based on various dimension reduction methods often make a further improvement over MARS. In particular, the improvement of the proposed drMARS is the most significant among the six SDR-MARS methods; see the columns under "SDR-MARS". For data.2 and data.3 where the training size is relatively larger, the conventional MARS performs similarly to RF whereas drMARS often makes remarkable improvement. The last three data of Table \ref{rMSPE-MCR} are results of classification. It can be seen that SDR-MARS based on various dimension reduction methods again outperforms MARS. drMARS has much more significant improvement in data.6 over MARS than the other methods.

\begin{figure}[h!]
		\centering
		\includegraphics[height=0.3\textheight,width=0.9\textwidth]{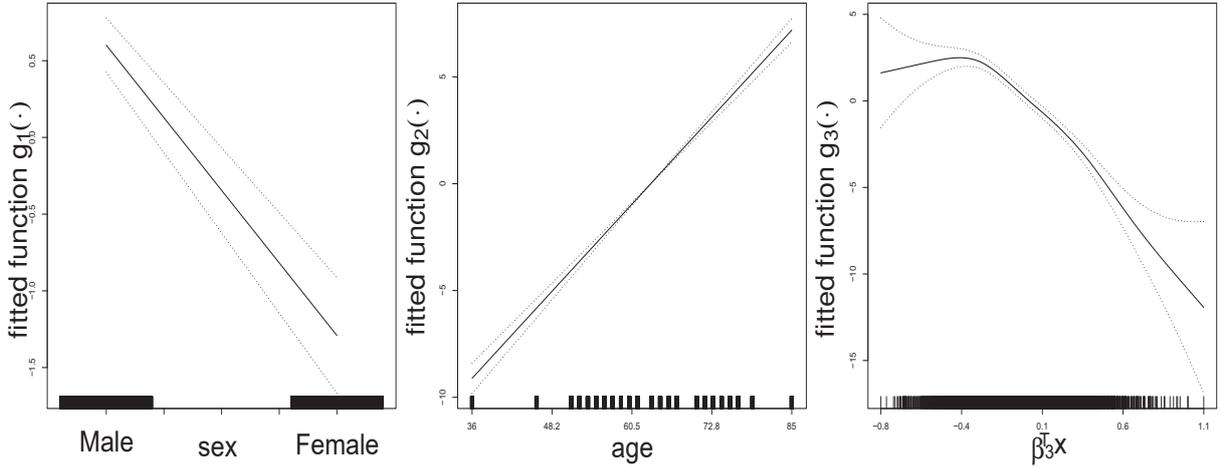}
		\vspace{-0cm}
		\caption{The estimated additive functions using the SDR directions for data.3}. \label{data3-xB}
	\end{figure}

	We next make some further illustration of the estimated model structure using data.3 and data.4. For ease of comparison, we standardize each variable. The estimation results of data.3 are listed in Table \ref{data3-dr}. The dimension of SDR space and the interaction degree of drMARS are selected as $3$ and $1$, respectively, and the regression model is estimated as follows,
	$$
	\begin{array}{rl}
		{\sf E}(Y\ |\ X=x) =&47.82+g_1(\beta_1^{^\intercal}x)+g_2(\beta_2^{^\intercal}x)+g_3(\beta_3^{^\intercal}x),\\
		g_1(v_1)=& 553.72(v_1-0.66)_+ +2.78(0.66-v_1)_+ +150.52(v_1-0.63)_+, \\
		g_2(v_2)=&-90.49(v_2+1.27)_+ -34.30(-1.27-v_2)_++121.43(v_2+0.00)_+\\ 
		&-111.66(v_2-0.19)_++28.26(v_2-0.57)_+  +12.22(v_2+0.41)_+\\
		& +158.01(v_2+1.01)_+ -123.24(v_2+0.65)_+,\\
		g_3(v_3)=& -7.67(v_3+0.29)_+ -8.60(-0.29-v_3)_+,
	\end{array}
	$$
	where ${\mathbf B} =(\beta_1, \beta_2, \beta_3)$ is the direction matrix in SDR space with the estimation results reported in Table \ref{data3-dr}. Note that the model is additive (with the interaction degree $1$), we are able to estimate each additive function (with confidence bands) as plotted in Figure \ref{data3-xB}.
	
	Note that as MARS takes a step-wise procedure to select the spline bases, some of them may be screened out. As a consequence, some of the predictors may not be selected in the model. This is shown clearly in the estimated coefficients of the directions in Table \ref{data3-dr}. It is known that gender and age are two important factors for the parkinson's disease, which are clearly dominant in the first two directions; see the first two columns of Table \ref{data3-dr} and the left and middle plots of Figure \ref{data3-xB}. It is also interesting to see that the last column in Table \ref{data3-dr} is mainly comprised of Shimmer.DDA (with coefficient 0.4893) and Jitter.RAP (with coefficient 0.4176) and Jitter.DDP (with coefficient -0.3865).  As the first two directions are mainly influenced by age and sex, we can plot them separately as shown in Figure \ref{data3-xB}, which is in line with the understanding of the relationship between the disease and the two variables. The estimated coefficients for the third projection imply that higher value of Shimmer.DDA leads to a lower degree of the disease. The result sheds some light on the debate about the usefulness of Shimmer.DDA in the disease diagnostics, suggesting that Shimmer.DDA is indeed useful in identifying the disease \citep[e.g.,][]{HJM07, KM14}.
	
	\begin{table}[h!]
	\centering
	\caption{The estimated SDR directions for data.3}\label{data3-dr}
			\begin{tabular}{lrrr}
				\toprule%
				X & $\beta_1$ &$\beta_2$ & $\beta_3$\\
				\midrule
				age   & -0.0237  & -0.9440  & 0.0722  \\
				sex   & -0.9443  & -0.0126  & 0.0399  \\
				test\_time & 0.0000  & 0.0000  & -0.0547  \\
				Jitter(\%)  & 0.0000  & 0.0000  & -0.1374  \\
				Jitter.Abs & 0.0000  & 0.0000  & 0.0606  \\
				Jitter.RAP & 0.0000  & 0.0106  & 0.4176  \\
				Jitter.PPQ5 & 0.0000  & 0.0000  & -0.1404  \\
				Jitter.DDP & 0.0000  & 0.0000  & -0.3865  \\
				Shimmer & 0.0000  & 0.0000  & -0.1577  \\
				Shimmer.dB & 0.0000  & 0.0000  & 0.2654  \\
				Shimmer.APQ3 & 0.0000  & 0.0000  & -0.3347  \\
				Shimmer.APQ5 & 0.0000  & 0.0000  & -0.1152  \\
				Shimmer.APQ11 & 0.0000  & 0.0000  & -0.0935  \\
				Shimmer.DDA & 0.0000  & 0.0126  & 0.4893  \\
				NHR   & 0.0000  & 0.0000  & 0.0635  \\
				HNR   & 0.0000  & 0.0000  & 0.0814  \\
				RPDE  & 0.0000  & 0.0000  & 0.0226  \\
				DFA   & 0.0000  & 0.0000  & 0.3361  \\
				PPE   & 0.0000  & 0.0000  & 0.0000  \\
				\bottomrule
			\end{tabular}
\end{table}
	

			For data.4, the dimension of SDR space and the interaction degree of MARS are selected as $3$ and $2$, respectively, and the model is estimated as follows,
			$$
			{\sf E}(Y\ |\ X=x) =5.82+g_1(\beta_1^{^\intercal}x)+g_2(\beta_2^{^\intercal}x)+g_3(\beta_3^{^\intercal}x)+g_{12}(\beta_1^{^\intercal}x,\beta_2^{^\intercal}x)+g_{13}(\beta_1^{^\intercal}x,\beta_3^{^\intercal}x),
			$$
			where $\beta_1 $,  $ \beta_2 $ and $ \beta_3$ are directions of the SDR space, and
			\begin{eqnarray*}
				g_1(v_1)&=&  31.67(v_1+0.00)_+ -42.83 (-0.00-v_1)_+\\
				                  &&-12.97(v_1+0.03)_+ -10.30 (v_1-0.03)_+ ,  \\
				g_2(v_2)&=& -7.52(v_2+0.00)_+ -3.73(-0.00-v_2)_+,\\
				g_3(v_3)&=& -28.42(v_2-0.03)_+ ,\\
				g_{12}(v_1,v_2)&=&881.25(-0.00-v_1)_+ (v_2-0.01)_+,\\
				g_{13}(v_1,v_3)&=&-153.80(v_1+0.03)_+ (v_3-0.01)_+.
		\end{eqnarray*}
	
		Inherited from MARS, drMARS may remove some variables if they are not important. Hence only a small portion of the 102 predictors are selected by drMARS in the final model and have nonzero coefficients in the directions, and the coefficients of the remaining variables are zero. Table \ref{data4-dr} only lists those variables that have nonzero coefficients. Note that the 102 predictors include 7 project physical and financial variables, and 5 groups of time lag economic variables (5*19 variables in total), which we denote in Table \ref{data4-dr} as lag $k$, $k= 1, \cdots, 5$.  It shows that none of project physical and financial variables is significant, and significant economic variables appear in multiple time lags, indicating that economic variables have a durable effect on final costs. Specifically,  the building services index ($x_9, x_{28}, x_{47}, x_{66}$) and consumer price index ($x_{22}, x_{23}, x_{41}, x_{42}, x_{60}, x_{61},$) are important factors for the final cost. The land price index ($x_{33}, x_{52}$) and the cumulative liquidity ($x_{12}, x_{31}, x_{50}, x_{69}$) also affect the final cost.
		
		\begin{table}[h!]
			\centering
			\caption{The estimated SDR directions for data.4 with nonzero coefficients, while coefficients for those not listed here are all 0.}\label{data4-dr}
				\setlength{\tabcolsep}{0.1mm}{
					\begin{tabular}{lrrr}
						\toprule%
						variables that has non-zero coefficients in drMARS & $\beta_1$ &$\beta_2$ & $\beta_3$\\
						\midrule
						$x_9$: Building services index (BSI) for a preselected base year (lag 1) & 0.4291  & -0.2057  & 0.1172  \\
						$x_{12}$: Cumulative liquidity (lag 1) & 0.1303  & 0.0000  & 0.1469  \\
						$x_{22}$: Consumer price index (CPI) in the base year (lag 1) & 0.0000  & -0.1739  & 0.0000  \\
						$x_{23}$: CPI of housing, water, fuel \& power in the base year (lag 1) & 0.3816  & 0.2211  & 0.0000  \\
						$x_{28}$: Building services index (BSI) for a preselected base year (lag 2) & -0.5875  & 0.0000  & 0.0000  \\
						$x_{29}$: Wholesale price index (WPI) of building materials for the base  & \multicolumn{1}{c}{\multirow{2}[0]{*}{0.0000  }} & \multicolumn{1}{c}{\multirow{2}[0]{*}{-0.1322  }} & \multicolumn{1}{c}{\multirow{2}[0]{*}{0.0000 }} \\
						\ \ \ \ \ \ \ \ year (lag 2) &       &       &  \\
						$x_{31}$: Cumulative liquidity (lag 2) & 0.0000  & -0.1167  & -0.1606  \\
						$x_{33}$: Land price index for the base year (lag 2) & 0.0000  & 0.1473  & 0.0000  \\
						$x_{41}$: Consumer price index (CPI) in the base year (lag 2) & 0.2372  & 0.0000  & 0.1360  \\
						$x_{42}$: CPI of housing, water, fuel \& power in the base year (lag 2) & -0.2168  & -0.5692  & -0.3176  \\
						$x_{47}$: Building services index (BSI) for a preselected base year (lag 3) & 0.1970  & 0.6146  & 0.3905  \\
						$x_{50}$: Cumulative liquidity (lag 3) & -0.1275  & 0.0000  & -0.3034  \\
						$x_{52}$: Land price index for the base year (lag 3) & -0.1467  & 0.0000  & -0.1483  \\
						$x_{60}$: Consumer price index (CPI) in the base year (lag 3) & 0.0000  & -0.1859  & 0.0000  \\
						$x_{61}$: CPI of housing, water, fuel \& power in the base year (lag 3) & 0.1178  & 0.2444  & 0.4858  \\
						$x_{66}$: Building services index (BSI) for a preselected base year (lag 4) & -0.2091  & 0.0000  & -0.4764  \\
						$x_{69}$: Cumulative liquidity (lag 4) & 0.1549  & 0.0000  & 0.2307  \\
						\bottomrule
					\end{tabular}}
		\end{table}
	
			\begin{figure}[h!]
			\centering
			\includegraphics[height=0.25\textheight,width=\textwidth]{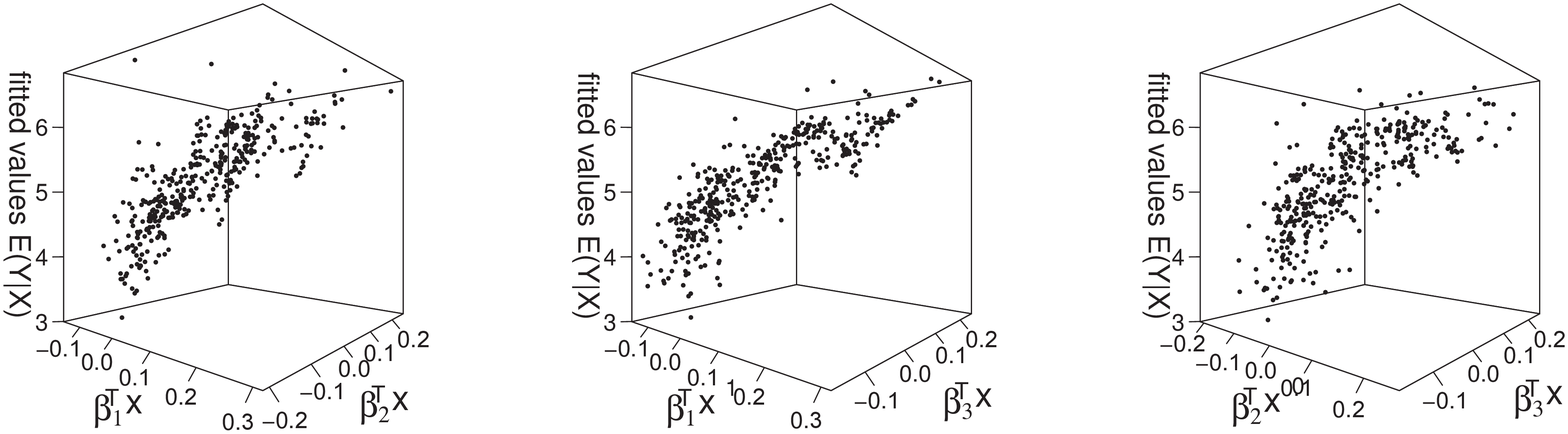}
			\vspace{-0.1cm}
			\includegraphics[height=0.25\textheight,width=\textwidth]{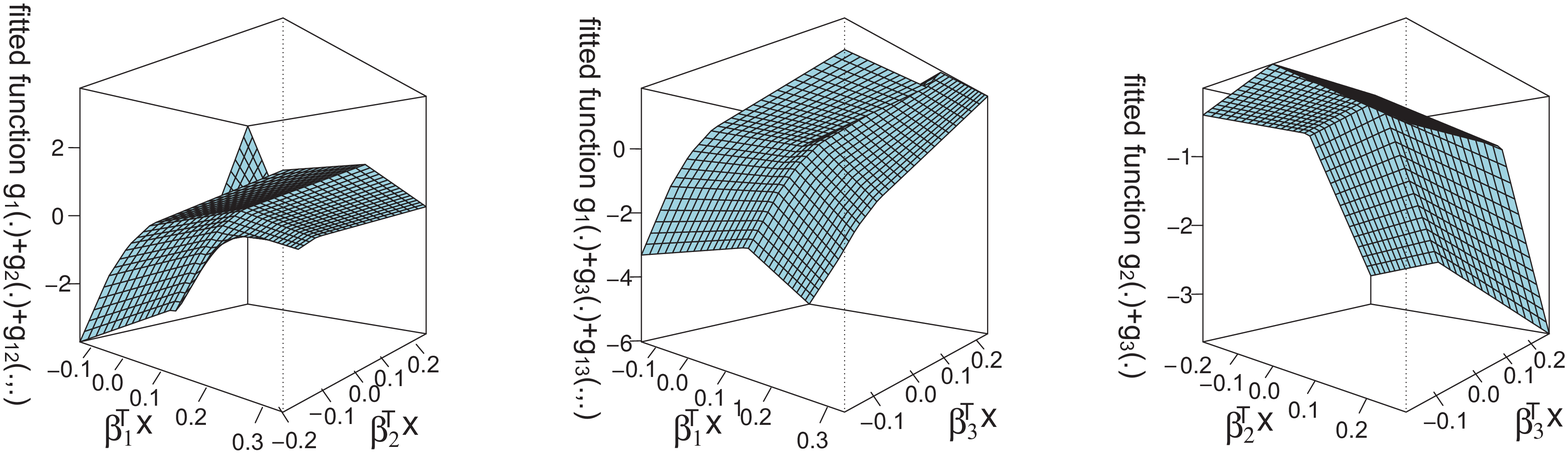}
			\vspace{-0.3cm}
			\caption{Plots for data.4: the plots on the top panel are the cost against each direction, and those on the bottom panel are the fitted functions}. \label{data4-xB}
		\end{figure}
	
		Due to the interaction degree $2$, we draw plots in the three-dimensional space for the dependence of the cost $ Y $ on the projected variables $\beta_j^{^\intercal}X, j=1, 2, 3$, as shown  in Figure \ref{data4-xB}. The first row of plots shows the relationship between cost and the directions; the second row shows the corresponding fitted functions specified in the estimated model above. Interestingly, as shown in the last panel of Figure \ref{data4-xB}, the cost has a non-linear dependence on the direction. The nonlinearity is along the second direction, which is a contrast between BSI for a preselected base year $ x_{47} $ (coefficient 0.6146) with the CPI of housing, water, fuel \& power in the base year $x_{42}, x_{61} $ (coefficient -0.5692, 0.2444). The reason for this nonlinearity needs further investigation.
		
		
		\section{Conclusion}\label{sec6}
		
		This paper has proposed a general method that combines SDR with the commonly-used MARS algorithm to estimate nonparametric regression functions. The special structure of the MARS basis functions makes it easy to compute the gradient vector of regression functions and thus the SDR space. The selection of spline functions in MARS also makes our dimension reduction method more suitable for high dimensional data. The proposed drMARS based on the SDR space can in turn improve the efficiency of conventional MARS. Through the comparison with other commonly-used nonparametric estimation and dimension reduction techniques, our numerical studies including both simulation and empirical applications show that the proposed drMARS has better finite-sample performance in both in-sample estimation and out-of-sample prediction.
		
		Several issues can be studied further. \cite{Xia2022} suggest a hybrid of random projection with SDR, which uses random projection to reduce the dimension of predictors to a lower-dimensional space and then applies SDR to the smaller space. We conjecture that such a hybrid may be adopted here. Our method can also be applied to other regression methods such as the random forest or support vector machine to solve the interaction between variables.
		
		




\appendix
\section*{Appendix: Proofs of the asymptotic theorems}
\label{app:theorem}



In this appendix we prove the main theorems in Sections \ref{sec2} and \ref{sec3}. Throughout the proofs, we let $C$ denote a generic positive constant whose value may change from line to line. We start with a useful inequality for independent random matrices \citep{Tropp12}.

\begin{lemma}\label{le:1}

Suppose that ${\boldsymbol\Lambda}_i$, $i=1,\cdots,n$, are independent $q_1\times q_2$ random matrices with zero mean and $\max_{1\leq i\leq n}\Vert{\boldsymbol\Lambda}_i\Vert\leq\lambda_n$. Then for any $z>0$,
\[
{\sf P}\left(\left\Vert \sum_{i=1}^n{\boldsymbol\Lambda}_i\right\Vert\geq z\right)\leq (q_1+q_2)\exp\left\{-\frac{z^2}{2(\xi_n^2+\lambda_nz/3)}\right\},
\]
where 
\[
\xi_n^2=\max\left\{\left\Vert\sum_{i=1}^n {\sf E}\left[{\boldsymbol\Lambda}_i{\boldsymbol\Lambda}_i^{^\intercal}\right]\right\Vert, \left\Vert\sum_{i=1}^n {\sf E}\left[{\boldsymbol\Lambda}_i^{^\intercal}{\boldsymbol\Lambda}_i\right]\right\Vert\right\}.
\]

\end{lemma}

The following lemma ensures that the least squares estimate (\ref{est-LS}) is well defined.

\begin{lemma}\label{le:2}

Suppose that Assumption \ref{ass:1}(i)--(iii) is satisfied. Then $\frac{1}{n}\widetilde{\mathbb H}^{^\intercal}\widetilde{\mathbb H}$ is positive definite w.p.a.1.

\end{lemma}

\noindent{\bf Proof of Lemma \ref{le:2}}. Recall that ${\mathbb H}=\left[{\mathbf H}(X_1),\cdots, {\mathbf H}(X_n)\right]^{^\intercal}$. We first prove
\begin{equation}\label{eqA.1}
\left\Vert \frac{1}{n}{\mathbb H}^{^\intercal}{\mathbb H}-{\boldsymbol\Omega}\right\Vert=o_P(1).
\end{equation}

Note that
\[
\frac{1}{n}{\mathbb H}^{^\intercal}{\mathbb H}-{\boldsymbol\Omega}=\frac{1}{n}\sum_{i=1}^n\left[{\mathbf H}(X_i){\mathbf H}(X_i)^{^\intercal}-{\boldsymbol\Omega}\right].
\]
We next make use of the inequality in Lemma \ref{le:1} with ${\boldsymbol\Lambda}_i={\mathbf H}(X_i){\mathbf H}(X_i)^{^\intercal}-{\boldsymbol\Omega}$ to prove (\ref{eqA.1}). It is easy to verify that $\lambda_n=c_1m_H$ and $\xi_n^2=c_2m_H^2$, where $c_1$ and $c_2$ are two positive constants. 
By Lemma \ref{le:1} and $m_H\sqrt{\log m_H}=o(n)$, for any $\epsilon>0$, we have
\begin{eqnarray}
{\sf P}\left(\left\Vert \frac{1}{n}{\mathbb H}^{^\intercal}{\mathbb H}-{\boldsymbol\Omega}\right\Vert\geq\epsilon\right)&=&{\sf P}\left(\left\Vert \sum_{i=1}^n\left[{\mathbf H}(X_i){\mathbf H}(X_i)^{^\intercal}-{\boldsymbol\Omega}\right]\right\Vert\geq n\epsilon\right)\notag\\
&\leq&2m_H\exp\left\{-\frac{\epsilon^2 n^2}{2\left(c_2m_H^2+\epsilon c_1m_H n/3\right)}\right\}\notag\\
&\leq&2m_H\exp\left\{-\frac{\epsilon^2 n^2}{3c_2m_H^2}\right\}\notag\\
&=&\exp\left\{\log(2m_H)-\frac{\epsilon^2}{3c_2}\cdot\frac{n^2}{m_H^2}\right\}=o(1),\notag
\end{eqnarray}
completing the proof of (\ref{eqA.1}). 

Combining Assumption \ref{ass:1}(iii) with (\ref{eqA.1}), we may show that $\frac{1}{n}{\mathbb H}^{^\intercal}{\mathbb H}$ is positive definite w.p.a.1, i.e., $\lambda_{\min}(\frac{1}{n}{\mathbb H}^{^\intercal}{\mathbb H})$ is positive and bounded away from zero, where $\lambda_{\min}(\cdot)$ denotes the minimum eigenvalue of a square matrix. It is trivial to verify that 
\[\lambda_{\min}({\mathbb H}^{^\intercal}{\mathbb H})
\leq \lambda_{\min}(\widetilde{\mathbb H}^{^\intercal}\widetilde{\mathbb H}).
\]
Hence, we may claim that $\frac{1}{n}\widetilde{\mathbb H}^{^\intercal}\widetilde{\mathbb H}$ is positive definite w.p.a.1. \hfill$\Box$\\

\bigskip

\noindent{\bf Proof of Theorem \ref{th:1}}. Without loss of generality, we next prove the convergence results by setting $\widetilde m=m_\ast$, where $m_\ast$ is a non-random positive integer. Letting ${\boldsymbol\varepsilon}=(\varepsilon_1,\cdots,\varepsilon_n)^{^\intercal}$ and ${\mathbf G}=\left[G(X_1),\cdots,G(X_n)\right]^{^\intercal}$, by (\ref{est-LS}), we have
\begin{equation}\label{eqA.2}
	\widetilde{G}^\prime(x)-G^\prime(x)=\Pi_1(x)+\Pi_2(x),
\end{equation}
where
\[\Pi_1(x)=\widetilde{\mathbf H}^\prime(x)^{^\intercal}\left(\widetilde{\mathbb H}^{^\intercal}\widetilde{\mathbb H}\right)^{-1}\widetilde{\mathbb H}^{^\intercal} {\boldsymbol\varepsilon},\ \ \Pi_2(x)=\widetilde{\mathbf H}^\prime(x)^{^\intercal}\left(\widetilde{\mathbb H}^{^\intercal}\widetilde{\mathbb H}\right)^{-1}\widetilde{\mathbb H}^{^\intercal} {\mathbf G}-G^\prime(x).\]				

We first consider $\Pi_1(x)$. Let ${\cal F}_X=\sigma(X_1,\cdots,X_n)$ and ${\mathbf E}_{m_\ast}=\left(e_1, e_2, \cdots, e_{m_\ast}\right)^{^\intercal}$, where $e_{j}$ is an $m_H$-dimensional vector with the $j$-th element being 1 and the others being zeros. Note that $\widetilde{\mathbb H}^{^\intercal}={\mathbf E}_{m_\ast}{\mathbb H}^{^\intercal}$. By Assumption \ref{ass:1}(i), we have
\begin{equation}\label{eqA.3}
	{\mathbf E}_{m_\ast}{\sf Var}\left(n^{-1/2}{\mathbb H}^{^\intercal} {\boldsymbol\varepsilon}\ |\ {\cal F}_X\right){\mathbf E}_{m_\ast}^{^\intercal}=\sigma^2 \left(\frac{1}{n}{\mathbf E}_{m_\ast}{\mathbb H}^{^\intercal}{\mathbb H}{\mathbf E}_{m_\ast}^{^\intercal}\right)=\sigma^2 \left(\frac{1}{n}\widetilde{\mathbb H}^{^\intercal}\widetilde{\mathbb H}\right),
\end{equation}
indicating that 
\begin{equation}\label{eqA.4}
	\vert\Pi_1(x)\vert_2^2\leq \frac{C}{n}\cdot \left\Vert\widetilde{\mathbf H}^\prime(x)^{^\intercal}\left(\frac{1}{n}\widetilde{\mathbb H}^{^\intercal}\widetilde{\mathbb H}\right)^{-1}\widetilde{\mathbf H}^\prime(x)\right\Vert\ \ {\rm w.p.a.1.}
\end{equation}
As $\Vert\widetilde{\mathbf H}^\prime(x)\Vert$ is of order $m_\ast^{1/2}$, it follows from (\ref{eqA.4}) and Lemma \ref{le:2} that
\begin{equation}\label{eqA.5}
	\left\vert \Pi_1(x) \right\vert_2=O_P\left(m_\ast^{1/2}n^{-1/2}\right).
\end{equation}
On the other hand, by Assumption \ref{ass:1}(iv), we have
\begin{equation}\label{eqA.6}
	\left\vert\Pi_2(x)\right\vert_2=\left\vert \widetilde{\mathbf H}^\prime(x)^{^\intercal}\left(\widetilde{\mathbb H}^{^\intercal}\widetilde{\mathbb H}\right)^{-1}\widetilde{\mathbb H}^{^\intercal} {\mathbf G}-G^\prime(x)\right\vert_2=\left\vert \widetilde{\mathbf H}^\prime(x)^{^\intercal}{\boldsymbol\alpha}_o-G^\prime(x)\right\vert_2=O_P\left(\widetilde\rho(m_\ast)\right).
\end{equation}
By (\ref{eqA.5}) and (\ref{eqA.6}), we complete the proof of (\ref{th1-1}).

We next turn to the proof of (\ref{th1-2}). Note that
\begin{eqnarray}
	\widetilde{\boldsymbol\Sigma}_G-{\boldsymbol\Sigma}_G&=&\left[\frac{1}{n} \sum_{i=1}^n\widetilde{G}^\prime(X_i)\widetilde{G}^\prime(X_i)^{^\intercal}-\frac{1}{n} \sum_{i=1}^nG^\prime(X_i)G^\prime(X_i)^{^\intercal}\right]+\nonumber\\
	&&\left[\frac{1}{n} \sum_{i=1}^nG^\prime(X_i)G^\prime(X_i)^{^\intercal}-{\boldsymbol\Sigma}_G\right]\nonumber\\
	&=:&\Pi_3+\Pi_4.\label{eqA.7}
\end{eqnarray}
By Assumption \ref{ass:1}(i) and Lemma \ref{le:1}, we readily have
\[
{\sf P}\left(\left\Vert\frac{1}{n} \sum_{i=1}^nG^\prime(X_i)G^\prime(X_i)^{^\intercal}-{\boldsymbol\Sigma}_G\right\Vert\geq M n^{-1/2}\right)\rightarrow0
\]
when $M\rightarrow\infty$. This indicates that
\begin{equation}\label{eqA.8}
	\Vert \Pi_4\Vert=\left\Vert\frac{1}{n} \sum_{i=1}^nG^\prime(X_i)G^\prime(X_i)^{^\intercal}-{\boldsymbol\Sigma}_G\right\Vert=O_P\left(n^{-1/2}\right).
\end{equation}
Re-write $\Pi_3$ as
\begin{eqnarray}
	\Pi_3&=&\frac{1}{n} \sum_{i=1}^n\left[\widetilde{G}^\prime(X_i)-G^\prime(X_i)\right]\left[G^\prime(X_i)\right]^{^\intercal}+\frac{1}{n} \sum_{i=1}^n\left[G^\prime(X_i)\right]\left[\widetilde G^\prime(X_i)-G^\prime(X_i)\right]^{^\intercal}+\nonumber\\
	&&\frac{1}{n} \sum_{i=1}^n\left[\widetilde{G}^\prime(X_i)-G^\prime(X_i)\right]\left[\widetilde{G}^\prime(X_i)-G^\prime(X_i)\right]^{^\intercal}\nonumber\\
	&=:&\Pi_{3,1}+\Pi_{3,2}+\Pi_{3,3}.\label{eqA.9}
\end{eqnarray}
Following the proofs of (\ref{eqA.3})--(\ref{eqA.6}), we may show that 
\begin{equation}\label{eqA.10}
\frac{1}{n}\sum_{i=1}^n\vert\Pi_1(X_i)\vert_2\leq Cn^{-1/2}\cdot \frac{1}{n}\sum_{i=1}^n\left\Vert\widetilde{\mathbf H}^\prime(X_i)\right\Vert= O_P\left(n^{-1/2}m_\ast^{1/2}\right),
\end{equation}
and 
\begin{equation}\label{eqA.11}
\frac{1}{n}\sum_{i=1}^n\vert\Pi_2(X_i)\vert_2=O_P\left(\widetilde\rho(m_\ast)\right).  
\end{equation}
By the decomposition (\ref{eqA.2}), the Cauchy-Schwarz inequality, (\ref{eqA.10}) and (\ref{eqA.11}), we have
\begin{eqnarray}
	\Vert \Pi_{3,1}\Vert&\leq&\frac{1}{n} \sum_{i=1}^n\left\Vert\left[\widetilde{G}^\prime(X_i)-G^\prime(X_i)\right]\left[G^\prime(X_i)\right]^{^\intercal}\right\Vert\notag\\
	&\leq&\frac{1}{n}\sum_{i=1}^n\left\vert\widetilde{G}^\prime(X_i)-G^\prime(X_i)\right\vert_2 \left\vert G^\prime(X_i)\right\vert_2\notag\\
    &\leq&C\cdot\frac{1}{n}\sum_{i=1}^n\left\vert\widetilde{G}^\prime(X_i)-G^\prime(X_i)\right\vert_2 \notag\\
    &\leq&C\left(\frac{1}{n}\sum_{i=1}^n \vert\Pi_1(X_i)\vert_2+\frac{1}{n}\sum_{i=1}^n \vert\Pi_2(X_i)\vert_2\right) \notag\\
&=&O_P\left(m_\ast^{1/2}n^{-1/2}+\widetilde\rho(m_\ast)\right),\label{eqA.12}
\end{eqnarray}
and similarly
\begin{equation}\label{eqA.13}
	\Vert \Pi_{3,2}\Vert=O_P\left(m_\ast^{1/2}n^{-1/2}+\widetilde\rho(m_\ast)\right),\ \ \Vert \Pi_{3,3}\Vert=O_P\left(m_\ast n^{-1}+\widetilde\rho^2(m_\ast)\right).
\end{equation}
By virtue of (\ref{eqA.7})--(\ref{eqA.9}), (\ref{eqA.12}) and (\ref{eqA.13}), we complete the proof of (\ref{th1-2}).\hfill$\Box$\\

\bigskip

\noindent
{\bf Proof of Theorem \ref{th:2}}. By Assumption \ref{ass:1}(v) and the Davis-Kahan theorem \citep[e.g.,][]{YWS15}, there exists a $d\times d$ rotation matrix ${\mathbf Q}$ such that
\[
\left\Vert\widetilde{\mathbf B}-{\mathbf B}{\mathbf Q}\right\Vert\leq C\left\Vert\widetilde{\boldsymbol\Sigma}_G-{\boldsymbol\Sigma}_G\right\Vert,
\]
which, together with (\ref{th1-2}), proves Theorem \ref{th:2}.	\hfill$\Box$\\

The following lemma ensures that the least squares estimate (\ref{drMARS-est-1}) is well defined.

\begin{lemma}\label{le:3}

Suppose that Assumptions \ref{ass:1} and \ref{ass:2}(i)--(iii) are satisfied. Then $\frac{1}{n}\widehat{\mathbb H}_\ast^{^\intercal}\widehat{\mathbb H}_\ast$ is positive definite w.p.a.1.

\end{lemma}

\noindent{\bf Proof of Lemma \ref{le:3}}. Recall that $\widehat{\mathbb H}_\ast=[\widehat{\mathbf H}(X_1^\ast),\cdots, \widehat{\mathbf H}(X_n^\ast)]^{^\intercal}$ and define $\widehat{\mathbb H}_\circ=[\widehat{\mathbf H}(X_1^\circ),\cdots,$ $\widehat{\mathbf H}(X_n^\circ)]^{^\intercal}$. By Theorem \ref{th:2}, the smoothness property of the basis functions and Assumption \ref{ass:2}(iii), we have
\begin{equation}\label{eqA.14}
	\left\Vert \frac{1}{n}\widehat{\mathbb H}_{\ast}^{^\intercal}\widehat{\mathbb H}_\ast-\frac{1}{n}\widehat{\mathbb H}_{\circ}^{^\intercal}\widehat{\mathbb H}_\circ\right\Vert=O_P\left(\widehat m \cdot (\widetilde m^{1/2}n^{-1/2}+\widetilde\rho(\widetilde m))\right)=o_P(1).
\end{equation}
By (\ref{eqA.14}), it is sufficient to show that $\frac{1}{n}\widetilde{\mathbb H}_\circ^{^\intercal}\widetilde{\mathbb H}_\circ$ is positive definite w.p.a.1. This can be proved by using Assumption \ref{ass:2}(ii) and following the proof of Lemma \ref{le:2}.\hfill$\Box$\\

\bigskip

\noindent{\bf Proof of Theorem \ref{th:3}}. Without loss of generality, we next prove the consistency property by setting $\widehat m=m_\circ$, where $m_\circ$ is a non-random positive integer. Let ${\boldsymbol\varepsilon}=(\varepsilon_1,\cdots,\varepsilon_n)^{^\intercal}$ and ${\mathbf G}=\left[G(X_1),\cdots,G(X_n)\right]^{^\intercal}$ as in the proof of Theorem \ref{th:2}. Note that
\begin{eqnarray}
\widehat{G}_0(x_\ast)- G_0(x_\ast)&=&\widehat{\mathbf H}(x_\ast)^{^\intercal}\left(\widehat{\mathbb H}_{\ast}^{^\intercal} \widehat{\mathbb H}_\ast\right)^{-1}\widehat{\mathbb H}_{\ast}^{^\intercal} {\boldsymbol\varepsilon}+\left[\widehat{\mathbf H}(x_\ast)^{^\intercal}\left(\widehat{\mathbb H}_{\ast}^{^\intercal} \widehat{\mathbb H}_\ast\right)^{-1}\widehat{\mathbb H}_{\ast}^{^\intercal} {\mathbf G}-G_0(x_\ast)\right]\notag\\
&=&\widehat{\mathbf H}(x_\ast)^{^\intercal}\left(\widehat{\mathbb H}_{\ast}^{^\intercal} \widehat{\mathbb H}_\ast\right)^{-1}\widehat{\mathbb H}_{\ast}^{^\intercal} {\boldsymbol\varepsilon}+O_P\left(\widehat\rho(m_\circ)\right)\label{eqA.15}
\end{eqnarray}
conditional on $\widehat m=m_\circ$.

Letting $\overline{\mathbb H}_\circ=\left[\overline{\mathbf H}(X_1^\circ),\cdots, \overline{\mathbf H}(X_n^\circ)\right]^{^\intercal}$ with $\overline{\mathbf H}(\cdot)$ defined in Section \ref{sec3}, and ${\mathbf E}_{m_\circ}=\left(e_1, e_2, \cdots, e_{m_\circ}\right)^{^\intercal}$, we then have $\widehat{\mathbb H}_\circ^{^\intercal}={\mathbf E}_{m_\circ}\overline{\mathbb H}_\circ^{^\intercal}$. Note that
\begin{equation}\label{eqA.16}
	\widehat{\mathbb H}_{\ast}^{^\intercal} {\boldsymbol\varepsilon}=\widehat{\mathbb H}_{\circ}^{^\intercal}{\boldsymbol\varepsilon}+\left(\widehat{\mathbb H}_{\ast}-\widehat{\mathbb H}_{\circ}\right)^{^\intercal}{\boldsymbol\varepsilon}={\mathbf E}_{m_\circ}\overline{\mathbb H}_\circ^{^\intercal}{\boldsymbol\varepsilon}+\left(\widehat{\mathbb H}_{\ast}-\widehat{\mathbb H}_{\circ}\right)^{^\intercal}{\boldsymbol\varepsilon}.
\end{equation}
As in the proof of (\ref{eqA.3}), we may show that
\[
{\mathbf E}_{m_\circ}{\sf Var}\left(n^{-1/2}\overline{\mathbb H}_\circ^{^\intercal} {\boldsymbol\varepsilon}\ |\ {\cal F}_X\right){\mathbf E}_{m_\circ}^{^\intercal}=\sigma^2 \left(\frac{1}{n}{\mathbf E}_{m_\circ}\overline{\mathbb H}_\circ^{^\intercal}\overline{\mathbb H}_\circ{\mathbf E}_{m_\circ}^{^\intercal}\right)=\sigma^2 \left(\frac{1}{n}\widehat{\mathbb H}_\circ^{^\intercal}\widehat{\mathbb H}_\circ\right),
\]
which, together with Lemma \ref{le:3}, indicates that
\begin{eqnarray}
 \left\vert\widehat{\mathbf H}(x_\ast)^{^\intercal}\left(\widehat{\mathbb H}_{\ast}^{^\intercal} \widehat{\mathbb H}_\ast\right)^{-1}\widehat{\mathbb H}_{\circ}^{^\intercal} {\boldsymbol\varepsilon}\right\vert&=& \left\vert\widehat{\mathbf H}(x_\ast)^{^\intercal}\left(\widehat{\mathbb H}_{\circ}^{^\intercal} \widehat{\mathbb H}_\circ\right)^{-1}\widehat{\mathbb H}_{\circ}^{^\intercal} {\boldsymbol\varepsilon}\right\vert(1+o_P(1))\notag\\
 &=&O_P\left(m_\circ^{1/2}n^{-1/2}\right).  \label{eqA.17}   
\end{eqnarray}
On the other hand, by Theorem \ref{th:2}, Lemma \ref{le:3} and the smoothness property of the MARS basis functions, we have
\begin{equation}\label{eqA.18}
	\left\vert \widehat{\mathbf H}(x_\ast)^{^\intercal}\left(\widehat{\mathbb H}_{\ast}^{^\intercal} \widehat{\mathbb H}_\ast\right)^{-1}\left(\widehat{\mathbb H}_{\ast}-\widehat{\mathbb H}_{\circ}\right)^{^\intercal}{\boldsymbol\varepsilon}\right\vert=o_P\left(m_\circ^{1/2}n^{-1/2}\right).
\end{equation}
By virtue of (\ref{eqA.16})--(\ref{eqA.18}), we have
\begin{equation}\label{eqA.19}
	\left\vert \widehat{\mathbf H}(x_\ast)^{^\intercal}\left(\widehat{\mathbb H}_{\ast}^{^\intercal} \widehat{\mathbb H}_\ast\right)^{-1}\widehat{\mathbb H}_{\ast}^{^\intercal} {\boldsymbol\varepsilon}\right\vert=O_P\left(m_\circ^{1/2}n^{-1/2}\right)
\end{equation}
With (\ref{eqA.15}) and (\ref{eqA.19}), we complete the proof of (\ref{th3}).\hfill$\Box$\\


\section*{Acknowledgements}

The authors would like to thank an Editor and three reviewers for the constructive comments, which helped to improve the article. This project is supported by the National Natural Science Foundation of China (No. 72033002).

\vskip 0.2in
\bibliography{drMARSref}

\begin{thebibliography}{43}
\providecommand{\natexlab}[1]{#1}
\providecommand{\url}[1]{\texttt{#1}}
\expandafter\ifx\csname urlstyle\endcsname\relax
  \providecommand{\doi}[1]{doi: #1}\else
  \providecommand{\doi}{doi: \begingroup \urlstyle{rm}\Url}\fi

\bibitem[Bagnall et~al.(2018)Bagnall, Flynn, Large, Line, Bostrom, and
  Cawley]{bagnall2020rotation}
Anthony Bagnall, M~Flynn, J~Large, J~Line, A~Bostrom, and G~Cawley.
\newblock Is rotation forest the best classifier for problems with continuous
  features?
\newblock \emph{arXiv preprint arXiv:1809.06705}, 2018.

\bibitem[Bickel and Levina(2008)]{BL08}
Peter~J Bickel and Elizaveta Levina.
\newblock Covariance regularization by thresholding.
\newblock \emph{The Annals of Statistics}, 36\penalty0 (6):\penalty0
  2577--2604, 2008.

\bibitem[Blaser and Fryzlewicz(2016)]{BlaserFryzlewicz2016}
Rico Blaser and Piotr Fryzlewicz.
\newblock Random rotation ensembles.
\newblock \emph{Journal of Machine Learning Research}, 17\penalty0
  (1):\penalty0 126--151, 2016.

\bibitem[Bradley(2005)]{B05}
Richard~C Bradley.
\newblock Basic properties of strong mixing conditions. a survey and some open
  questions.
\newblock \emph{Probability Surveys}, 2:\penalty0 107--144, 2005.

\bibitem[Breiman(2001)]{breiman01}
Leo Breiman.
\newblock Random forests.
\newblock \emph{Machine Learning}, 45:\penalty0 5--32, 2001.

\bibitem[Cai et~al.(2022)Cai, Xia, and Hang]{Xia2022}
Zhibo Cai, Yingcun Xia, and Weiqiang Hang.
\newblock An outer-product-of-gradient approach to dimension reduction and its
  application to classification in high dimensional space.
\newblock \emph{Journal of the American Statistical Association}, forthcoming,
  2022.

\bibitem[Cannings and Samworth(2017)]{cannings2017random}
Timothy~I Cannings and Richard~J Samworth.
\newblock Random-projection ensemble classification.
\newblock \emph{Journal of the Royal Statistical Society: Series B (Statistical
  Methodology)}, 79\penalty0 (4):\penalty0 959--1035, 2017.

\bibitem[Chen et~al.(2010)Chen, Zou, and Cook]{chen2010coordinate}
Xin Chen, Changliang Zou, and Dennis Cook.
\newblock Coordinate-independent sparse sufficient dimension reduction and
  variable selection.
\newblock \emph{The Annals of Statistics}, 38\penalty0 (6):\penalty0
  3696--3723, 2010.

\bibitem[Cook and Li(2002)]{cook2002dimension}
Dennis Cook and Bing Li.
\newblock Dimension reduction for conditional mean in regression.
\newblock \emph{The Annals of Statistics}, 30\penalty0 (2):\penalty0 455--474,
  2002.

\bibitem[Cook and Li(2004)]{cook04}
R~Dennis Cook and Bing Li.
\newblock Determining the dimension of iterative hessian transformation.
\newblock \emph{The Annals of Statistics}, 32\penalty0 (6):\penalty0
  2501--2531, 2004.

\bibitem[Cortes and Vapnik(1995)]{cortes95}
Corinna Cortes and Vladimir Vapnik.
\newblock Support-vector networks.
\newblock \emph{Machine Learning}, 20:\penalty0 273--297, 1995.

\bibitem[Dimitriadou et~al.(2008)Dimitriadou, Hornik, Leisch, Meyer, and
  Weingessel]{dimitriadou08}
Evgenia Dimitriadou, Kurt Hornik, Friedrich Leisch, David Meyer, and Andreas
  Weingessel.
\newblock Misc functions of the department of statistics (e1071), tu wien.
\newblock \emph{R package}, 1:\penalty0 5--24, 2008.

\bibitem[Engle et~al.(1986)Engle, Granger, Rice, and Weiss]{EGRW86}
Robert~F. Engle, C.~W.~J. Granger, John Rice, and Andrew Weiss.
\newblock Semiparametric estimates of the relation between weather and
  electricity sales.
\newblock \emph{Journal of the American Statistical Association}, 81\penalty0
  (394), 1986.

\bibitem[Fan and Gijbels(1996)]{FG96}
Jianqing Fan and Irene Gijbels.
\newblock \emph{Local Polynomial Modelling and Its Applications}.
\newblock Chapman \& Hall/CRC, 1996.

\bibitem[Fertl and Bura(2022)]{fertl22}
Lukas Fertl and Efstathia Bura.
\newblock Conditional variance estimator for sufficient dimension reduction.
\newblock \emph{Bernoulli}, 28\penalty0 (3):\penalty0 1862--1891, 2022.

\bibitem[Friedman(1991)]{F91}
Jerome~H Friedman.
\newblock Multivariate adaptive regression splines.
\newblock \emph{The Annals of Statistics}, 19\penalty0 (1):\penalty0 1--67,
  1991.

\bibitem[Fukumizu and Leng(2014)]{FL14}
Kenji Fukumizu and Chenlei Leng.
\newblock Gradient-based kernel dimension reduction for regression.
\newblock \emph{Journal of the American Statistical Association}, 109\penalty0
  (505):\penalty0 359--370, 2014.

\bibitem[H\"ardle et~al.(1993)H\"ardle, Hall, and Ichimura]{HHI93}
Wolfgang H\"ardle, Peter Hall, and Hidehiko Ichimura.
\newblock Optimal smoothing in single-index models.
\newblock \emph{The Annals of Statistics}, 21\penalty0 (1):\penalty0 157--178,
  1993.

\bibitem[Hastie and Tibshirani(1986)]{HT86}
Trevor Hastie and Robert Tibshirani.
\newblock Generalized additive models.
\newblock \emph{Statistical Science}, 1\penalty0 (3):\penalty0 297--310, 1986.

\bibitem[Hastie and Tibshirani(1993)]{HT93}
Trevor Hastie and Robert Tibshirani.
\newblock Varying-coefficient models.
\newblock \emph{Journal of the Royal Statistical Society: Series B (Statistical
  Methodology)}, 55\penalty0 (4):\penalty0 757--796, 1993.

\bibitem[Hastie et~al.(2009)Hastie, Tibshirani, Friedman, and Friedman]{HTF09}
Trevor Hastie, Robert Tibshirani, Jerome~H Friedman, and Jerome~H Friedman.
\newblock \emph{The Elements of Statistical Learning: Data mining, Inference,
  and Prediction}, volume~2.
\newblock Springer, 2009.

\bibitem[Hausdorff(2007)]{HJM07}
Jeffrey~M Hausdorff.
\newblock Gait dynamics, fractals and falls: finding meaning in the
  stride-to-stride fluctuations of human walking.
\newblock \emph{Human Movement Science}, 26\penalty0 (4):\penalty0 555--589,
  2007.

\bibitem[Huang(2003)]{H03}
Jianhua~Z Huang.
\newblock Local asymptotics for polynomial spline regression.
\newblock \emph{The Annals of Statistics}, 31\penalty0 (5):\penalty0
  1600--1635, 2003.

\bibitem[Kirchner et~al.(2014)Kirchner, Schubert, Liebherr, and Haas]{KM14}
Marietta Kirchner, Patric Schubert, Magnus Liebherr, and Christian~T Haas.
\newblock Detrended fluctuation analysis and adaptive fractal analysis of
  stride time data in parkinson's disease: stitching together short gait
  trials.
\newblock \emph{PloS One}, 9\penalty0 (1):\penalty0 e85787, 2014.

\bibitem[Li(1991)]{L91}
Ker-Chau Li.
\newblock Sliced inverse regression for dimension reduction.
\newblock \emph{Journal of the American Statistical Association}, 86\penalty0
  (414):\penalty0 316--327, 1991.

\bibitem[Liaw and Wiener(2002)]{liaw15}
Andy Liaw and Matthew Wiener.
\newblock Classification and regression by randomforest.
\newblock \emph{R news}, 2\penalty0 (3):\penalty0 18--22, 2002.

\bibitem[Lin(2013)]{L13}
Wei Lin.
\newblock \emph{The Econometric Analysis of Interval-Valued Data and Adaptive
  Regression Splines}.
\newblock PhD thesis, UC Riverside, 2013.

\bibitem[Luo et~al.(2014)Luo, Li, and Yin]{luo2014efficient}
Wei Luo, Bing Li, and Xiangrong Yin.
\newblock On efficient dimension reduction with respect to a statistical
  functional of interest.
\newblock \emph{The Annals of Statistics}, 42\penalty0 (1):\penalty0 382--412,
  2014.

\bibitem[Ma and Zhu(2014)]{Ma14}
Yanyuan Ma and Liping Zhu.
\newblock On estimation efficiency of the central mean subspace.
\newblock \emph{Journal of the Royal Statistical Society: Series B (Statistical
  Methodology)}, 76\penalty0 (5):\penalty0 885--901, 2014.

\bibitem[Milborrow et~al.(2017)Milborrow, Hastie, Tibshirani, Miller, and
  Lumley]{earth23}
Stephen Milborrow, Trevor Hastie, Robert Tibshirani, Alan Miller, and Thomas
  Lumley.
\newblock earth: Multivariate adaptive regression splines.
\newblock \emph{R package version}, 5\penalty0 (2), 2017.

\bibitem[Stone(1982)]{S82}
Charles~J Stone.
\newblock Optimal global rates of convergence for nonparametric regression.
\newblock \emph{The Annals of Statistics}, 10\penalty0 (4):\penalty0
  1040--1053, 1982.

\bibitem[Stone(1990)]{S90}
Charles~J Stone.
\newblock Large-sample inference for log-spline models.
\newblock \emph{The Annals of Statistics}, 18\penalty0 (2):\penalty0 717--741,
  1990.

\bibitem[Stone(1991)]{S91}
Charles~J Stone.
\newblock Asymptotics for doubly flexible logspline response models.
\newblock \emph{The Annals of Statistics}, 19\penalty0 (4):\penalty0
  1832--1854, 1991.

\bibitem[Stone et~al.(1997)Stone, Hansen, Kooperberg, and Truong]{SHKT97}
Charles~J Stone, Mark~H. Hansen, Charles Kooperberg, and Young~K. Truong.
\newblock Polynomial splines and their tensor products in extended linear
  modeling.
\newblock \emph{The Annals of Statistics}, 25\penalty0 (4):\penalty0
  1371--1425, 1997.

\bibitem[Tropp(2012)]{Tropp12}
Joel~A. Tropp.
\newblock User-friendly tail bounds for sums of random matrices.
\newblock \emph{Foundations of Computational Mathematics}, 12:\penalty0
  389--434, 2012.

\bibitem[Wang et~al.(2015)Wang, Xu, and Zhu]{WXZ15}
Tao Wang, Peirong Xu, and Lixing Zhu.
\newblock Variable selection and estimation for semi-parametric multiple-index
  models.
\newblock \emph{Bernoulli}, 21\penalty0 (1):\penalty0 242--275, 2015.

\bibitem[Weisberg(2002)]{Weisberg02}
S.~Weisberg.
\newblock Dimension reduction regression in r.
\newblock \emph{Journal of Statistical Software}, 7\penalty0 (1):\penalty0
  1--22, 2002.

\bibitem[Xia(2008)]{X08}
Yingcun Xia.
\newblock A multiple-index model and dimension reduction.
\newblock \emph{Journal of the American Statistical Association}, 103\penalty0
  (484):\penalty0 1631--1640, 2008.

\bibitem[Xia et~al.(2002)Xia, Tong, Li, and Zhu]{XTLZ02}
Yingcun Xia, Howell Tong, Wai~Keung Li, and Li-Xing Zhu.
\newblock An adaptive estimation of dimension reduction space.
\newblock \emph{Journal of the Royal Statistical Society: Series B (Statistical
  Methodology)}, 64\penalty0 (3):\penalty0 363--410, 2002.

\bibitem[Yang et~al.(2017)Yang, Balasubramanian, and Liu]{YBL17}
Zhuoran Yang, Krishnakumar Balasubramanian, and Han Liu.
\newblock On stein's identity and near-optimal estimation in high-dimensional
  index models.
\newblock \emph{arXiv preprint arXiv:1709.08795}, 2017.

\bibitem[Yin and Li(2011)]{YL11}
Xiangrong Yin and Bing Li.
\newblock Sufficient dimension reduction based on an ensemble of minimum
  average variance estimators.
\newblock \emph{The Annals of Statistics}, 39\penalty0 (6):\penalty0
  3392--3416, 2011.

\bibitem[Yu et~al.(2015)Yu, Wang, and Samworth]{YWS15}
Yi~Yu, Tengyao Wang, and Richard~J Samworth.
\newblock A useful variant of the davis--kahan theorem for statisticians.
\newblock \emph{Biometrika}, 102\penalty0 (2):\penalty0 315--323, 2015.

\bibitem[Zhou et~al.(1998)Zhou, Shen, and Wolfe]{ZSW98}
S.~Zhou, X.~Shen, and D.A. Wolfe.
\newblock Local asymptotics for regression splines and confidence regions.
\newblock \emph{The Annals of Statistics}, 26\penalty0 (5):\penalty0
  1760--1782, 1998.

\end{thebibliography}

\end{document}